\documentclass[onecolumn, preprint]{aastex631}

\received{18 June 2024}
\revised{-}
\accepted{-}

\submitjournal{ApJ}

\graphicspath{{./}{figures/}}

\usepackage{physics, nicefrac, dutchcal}

\newcommand{\g}{g^{\nicefrac{1}{2}}}

\newcommand{\n}[1]{\textcolor{red}{note: #1}}

\begin{document}

\title{Distorted Magnetic Flux Ropes within Interplanetary Coronal Mass Ejections}

\author[0000-0002-6273-4320]{Andreas J. Weiss}
\affiliation{NASA Postdoctoral Program Fellowship, NASA Goddard Space Flight Center, Greenbelt, MD 20771, USA}
\affiliation{Heliophysics Science Division, NASA Goddard Space Flight Center, Greenbelt, MD 20771, USA}
\correspondingauthor{Andreas J. Weiss}\email{ajefweiss@gmail.com}

\author[0000-0003-0565-4890]{Teresa Nieves-Chinchilla}
\affiliation{Heliophysics Science Division, NASA Goddard Space Flight Center, Greenbelt, MD 20771, USA}

\author[0000-0001-6868-4152]{Christian M\"ostl}
\affiliation{Austrian Space Weather Office, GeoSphere Austria, Graz, Austria}

\begin{abstract}
Magnetic flux ropes within interplanetary coronal mass ejections are often characterized as simplistic cylindrical or toroidal tubes with field lines that twist around the cylinder or torus axis. Recent multi-point observations suggest that the overall geometry of these large-scale structures may be significantly more complex, so that the contemporary modeling approaches would be, in some cases, insufficient to properly understand the global structure of any interplanetary coronal mass ejection. In an attempt to partially rectify this issue, we have developed a novel magnetic flux rope model that allows for the description of arbitrary distortions of the cross-section or deformation of the magnetic axis. The distorted magnetic flux rope model is a fully analytic flux rope model, that can be used to describe significantly more complex geometries and is numerically efficient enough to be used for large ensemble simulations. To demonstrate the usefulness of our model, we focus on a specific implementation of our model and apply it to an ICME event that was observed \textit{in situ} on 2023 April 23 at the L1 point by the Wind spacecraft and also by the STEREO-A spacecraft that was $10.2^\circ$ further east and $0.9^\circ$ south in heliographic coordinates. We demonstrate that our model can accurately reconstruct each observation individual and also gives a fair reconstruction of both events simultaneously using a multi-point reconstruction approach, which results in a geometry that is not fully constistent with a cylindrical or toroidal approximation. 
\end{abstract}


\section{Introduction} \label{sec:intro}

A magnetic flux rope can be broadly defined as a flux tube with magnetic field lines that twist around a single common axis. Such magnetic field structures are known play a prominent role, and are found, in a many different solar or general astrophysical settings such as coronal mass ejections \citep[e.g.][]{vourlidas2014}, coronal loops \citep[e.g.][]{titov1999,wang2019}, flux transfer events \citep[e.g.][]{fargette2020,jasinski2021}, magnetotail plasma sheets \citep[e.g.][]{slavin2003a} or small scale magnetic flux ropes \citep[e.g.][]{moldwin2000}. For the purpose of this manuscript, we will be primarily focusing on magnetic flux ropes (MFRs) as they are understood within heliophysics community and used to describe the core region of interplanetary coronal mass ejections (ICMEs). Despite the fact that typical flux rope signatures are not necessarily always observed, due to what are thought to be primarily geometric effects \citep{song2020}, it is commonly thought that all ICMEs carry such a flux rope structure within them. Nonetheless, such ICME flux ropes are regularly observed \textit{in situ}, using magnetic field measurements provided by magnetometers, from spacecraft at various different distances from the Sun, such as Parker Solar Probe \citep{salman2024}, Solar Orbiter, or the multitude of existing L1 monitors \citep{richardson_data}. More recently, given the larger number of spacecraft missions within our heliosphere, certain single events have also been observed at multiple different locations \citep[e.g.][]{moestl2022}.  Particularly well-behaved events with clear signatures of a single rotating magnetic field component, these flux rope observations are sometimes also designated as magnetic clouds \citep[e.g.][]{burlaga1981,bothmer1998}.

One can attempt to describe flux rope observations using a variety of different idealized models \citep[e.g.][]{lepping1990,demoulin2019}. The two most common of these, are the linear force-free model \citep{lundquist1950} and the force-free uniform twist model \citep{gold1960} that are sometimes designated by the name of their respective authors. They assume the geometry of the flux rope structure to be an infinitely long cylinder with a force-free magnetic field, where the magnetic field lines twist around the cylinder axis. But unfortunately, many observations do not conform to this simplified picture, so that these idealized models cannot always be appropriately applied. The next level of model complexity can be achieved by including curvature and describing the flux rope as a torus \citep[e.g.][]{titov1999,vandas2017b}. These torus models allow for more asymmetric magnetic field profiles and a description of flank encounters, where the spacecraft never crosses over the magnetic axis, which is a scenario that is not well captured by a circular cylindrical geometry \citep[e.g.][]{marubashi2007}{}{}.

From remote sensing observations \citep[e.g.][]{deforest2013,davies2021}, magneto-hydrodynamic (MHD) simulations \citep[e.g.][]{toeroek2005}{}, and  kinematic studies \citep[e.g.][]{riley2004,owens2006} it is known that these cylindrical or torus models are strong approximations of reality, specifically when it comes to the circular shape of their cross-sections.
As such, more recent modeling efforts focus on using cylindrical or toroidal configurations with non-circular cross-sections \citep[e.g.][]{vandas2017a,nieves2018b,nieves2023}. These models attempt to mimic the distortions that are believed to occur due to the interaction of the ICME with the coronal magnetic field and the downstream solar wind environment, sometimes also referred to as the ``pancaking'' effect \citep[e.g.][]{titov1999}.
But even these more sophisticated models are unable to breech the gap between what the community believes that the overall geometry of an ICME or its flux rope actually look like and what can be modeled. Basic cartoon like figures that show the overall structure of ICMEs, such as those shown in \cite{zurbuchen2006}, describe a tube like structure that is attached on both ends to the Sun and expands significantly in size at the front section that is furthest away from the Sun. Variations of this concept are still in use today, and such sketches appear in almost every single peer reviewed paper that focuses on large-scale ICME topics. The existence of these problems has  led to the development of semi-empirical models, which attempt describe more complex geometries but which are either not fully analytical or sacrifice physical accuracy \citep[e.g.][]{isavnin2016}. These issues highlight a disconnect in between what the current overall understanding, or idea, on large-scale ICME or flux rope structure is and what tools are available to test the underlying assumptions. It can be expected that this particular topic will only become more relevant over the coming years due to the increase of multi-point ICME observations and future planned multi-spacecraft mission concepts \citep[e.g.][]{lugaz2023}

In this manuscript, we will attempt to rectify some of these problems by formulating a novel approach that allows the construction of almost arbitrarily complex ICME flux rope shapes that use  analytical expressions for describing both the geometry and the magnetic field. This model builds on top of recent developments that are showcased in \cite{weiss2022}, henceforth referred to as W22, and \cite{nieves2023}. We introduce the distorted magnetic flux rope (DMFR) in Section \ref{sec:model}, where we give a general description of the model in terms of arbitrary functions that characterize the overall geometry and magnetic field structure. We also briefly mention how we can find appropriate expressions for the magnetic field so that the model corresponds to the existing classical flux rope models so that one can make direct comparisons. For demonstration purposes, we have implemented the DMFR within the existing 3DCORE framework \citep[see][]{weiss2021b}, henceforth referred to as W21, so that we can make use of an existing Sequential Monte-Carlo algorithm to match our model with observations for specific events. An application of this implementation is shown in Section \ref{sec:appl}, where we match a multi-point event observed by the Wind and STEREO-A spacecraft with a specific global ICME flux rope shape and magnetic field configuration. The conclusions and a discussion of our overall results can be found in Section \ref{sec:conclusion}.

\section{Model} \label{sec:model}

Any magnetic flux rope model must describe a magnetic field that is solenoidal, so that the divergence $\div\vb*{B}$ vanishes everywhere. Introducing a general curvilinear coordinate system $\vb*{r}: [0, 1]^3 \mapsto \mathbb{R}^3$, with coordinates $(\mu,\,\nu,\,s)$, we can write down this condition as: 
\begin{linenomath*}
\begin{equation}\label{eq:solenoid}
0=\div\vb*{B}=\partial_\mu\left(\g B^\mu\right) + \partial_\nu\left(\g B^\nu\right) + \partial_s\left(\g B^s\right),
\end{equation}
\end{linenomath*}
where $g$ is the metric of the coordinate system and $B^i$ are the contravariant magnetic field components so \mbox{that $\vb*{B}=B^i\vb*{\epsilon}_i$}, where the basis vectors are given by $\vb*{\epsilon}_i=\partial_i\vb*{r}$. We designate the $\mu$ coordinate as the flux label, so that the surfaces implicitly defined by a constant $\mu$ coordinate describe flux surfaces to which the magnetic field lines are constrained. As such, the $\mu$ coordinate will be the analogue of the radial coordinate from a cylindrical coordinate and from the previous definition it automatically follows that the $B^\mu$ component must vanish everywhere. The $\nu$ and $s$ coordinates correspond to the poloidal and axial coordinate respectively, albeit in a highly obfuscated form. Such coordinates are commonly called flux coordinates \citep[see][]{d'haeseleer1991}, and assume that the magnetic field is configured in  a way that valid flux surfaces can be found. We will assume this to be the case, but it is a very strong assumption and in general there is no guarantee that such flux surfaces exist everywhere. This assumption precludes the ability to describe certain interesting magnetic field configurations, such as those that include magnetic islands or flux ropes with multiple magnetic axes. Flux coordinates are often used to describe a laboratory plasma within magnetic confinement devices, and examples of commonly used coordinate systems are the Boozer \citep[e.g.][]{boozer1981,rodriguez2021} or Hamada \citep[e.g.][]{hamada1962} coordinates. Within this context, the underlying challenge is to find an appropriate coordinate system and corresponding geometry that describes a plasma in an equilibrium state. For our case, these aspirations are partially irrelevant as the magnetic flux ropes that are found in the heliosphere can be considered to be evolving structures for which an equilibrium description is most likely not appropriate due to expansion \citep[e.g.][]{davies2021} and possible non-zero Lorentz forces \citep[e.g.][]{lynch2022}. Instead, we are primarily interested in being able to describe a complex geometry and magnetic field configuration that may be able explain the \textit{in situ} magnetic field measurements that we receive from spacecraft that are scattered throughout the heliosphere. As we will show, we can do this efficiently by using existing approaches with flux coordinates in combination with some of our previous results from W22. One downside in this new approach is that the expressions will look quite different from what is typically seen in any ICME related flux rope model, even when the magnetic field configuration is the same.


We start by finding an appropriate expression for the magnetic field that automatically satisfies the constraints given by Equation \eqref{eq:solenoid}. In any flux coordinate system, we can do this if we decompose the magnetic field in terms of our curvilinear basis vectors in the following way:
\begin{linenomath*}
\begin{equation}\label{eq:b_full}
\vb*{B} = B^\nu \vb*{\epsilon}_\nu + B^s \vb*{\epsilon}_s = \frac{\chi(\mu, s)+\partial_s \zeta(\mu,\nu,s)}{\g}\vb*{\epsilon}_\nu + \frac{\xi(\mu,\nu)-\partial_\nu \zeta(\mu,\nu,s)}{\g}\vb*{\epsilon}_s,
\end{equation}
\end{linenomath*}
where we introduce three new functions  $\chi(\mu, s),\ \xi(\mu, \nu)$ and $\zeta(\mu,\nu,s)$. These expressions have been written slightly different compared to some of the existing literature \citep[e.g.][]{d'haeseleer1991}, so that it more resembles what one would expect to see for flux rope models as they are used by heliophysics community. The first two of these largely determine the local twist of the magnetic field lines but are only dependent on two of three coordinates, which limits how the overall magnetic field can be configured. The third function allows us to include additional complexity, but only in an indirect way  as the function only appears in the magnetic field components in terms of a derivative. The above expressions will work for any flux rope geometry as long as it can be described in terms of flux coordinates, regardless of the curvature of the magnetic axis and no matter which cross-section shape is used. Because we are not interested in finding force-free or force balanced magnetic field configurations, we can make various different choices for $\chi,\xi$ or $\zeta$. Because the $g^{-\nicefrac{1}{2}}$ factor appears in both terms in Equation \eqref{eq:b_full}, the magnetic field will also depend on the specific choice of the coordinate system. As such, will the magnetic field will be significantly harder to configure for our purposes.


Note that this approach is now starkly different than the one used in W22, where we  mandated a specific expression for $B^s$ and then derived the appropriate corresponding expression for $B^\nu$ so that the magnetic field is divergence-less. While an extension of the approach used in W22 would in principle still work for an arbitrary coordinate system and geometry, we found that the resulting expressions were unwieldy. We briefly mention our efforts in that direction in Appendix \ref{app:w22_ext} as they still may be useful in certain specific scenarios. The advantage in this case, is that the resulting model will look significantly more similar to the standard force-free models that most people will be familiar with.

\subsection{Distorted Coordinates}
\label{sec:dtc}

Given our expression for the magnetic field components in Eq. \eqref{eq:b_full}, we now require a specific coordinate system to describe our flux rope geometry. For this purpose, we introduce the distorted coordinate system, that we define as:
\begin{linenomath*}
\begin{equation}
\vb*{r}(\mu,\nu,s)=\vb*{\gamma}(s)+\mathcal{D}(\mu,\nu,s)\Big[\vb*{\hat{n}}_1(s)\cos\Omega(\mu,\nu,s)+\vb*{\hat{n}}_2(s)\sin\Omega(\mu,\nu,s)\Big],\label{eq:coordinates}
\end{equation}
\end{linenomath*}
where $\vb*{\gamma}(s)$ is an arbitrarily parameterized space curve that follows the magnetic axis and the two functions $\mathcal{D}(\mu,\nu,s)$  and $\Omega(\mu,\nu,s)$ are the distortion and azimuth function, which both together will characterize the shape of the cross-section. 
The tangent vector to the curve $\vb*{\gamma}$ is given by $\hat{\vb*{t}}(s)=\partial_s\vb*{\gamma}/v(s)$, where $v(s)=\norm{\partial_s\vb*{\gamma}}$ is the curve velocity. By convention we assume that $\{\vb*{\hat{t}},\,\vb*{\hat{n}}_1,\vb*{\hat{n}}_2\}$ forms a right-handed orthogonal set. There are effectively two different options for defining the normal vectors $\vb*{n}_{1/2}$. The first approach, which at first glance appears to be simpler, is to use the Frenet-Serret frame that also neatly introduces the concept of curvature $\kappa(s)$ and torsion $\tau(s)$. The second alternative, which is also the only other option to describe the evolution of the normal vectors in terms of two independent quantities, is to use the approach that is introduced in \cite{bishop1975}. We called this approach the ``parallel transport frame'' in W22, and the normal vectors are implicitly defined by the following differential equations:
\begin{eqnarray}
\partial_s \vb*{\hat{t}} &=& v \left(k_1\,\vb*{\hat{n}}_1 + k_2\,\vb*{\hat{n}}_2\right),\label{eq:dtds}\\
\partial_s \vb*{\hat{n}}_1 &=& -v k_1\,\vb*{\hat{t}},\label{eq:dn1ds}\\
\partial_s \vb*{\hat{n}}_2 &=& -v k_2\,\vb*{\hat{t}},\label{eq:dn2ds}
\end{eqnarray}
where we introduce two curvature values $k_1(s)$ and $k_2(s)$. These curvature values are related to the curvature $\kappa$ via the relation $\kappa^2=k^2_1 + k^2_2$. Solving for Eqs. (\ref{eq:dtds}-\ref{eq:dn2ds}) also requires choosing initial conditions for the normal vectors, so that the solution is not unique.  Using the parallel transport frame, to generate the normal vectors $\vb*{\hat{n}}_{1/2}$, we can compute the basis vectors of our coordinate system that take on the following form:
\begin{eqnarray}
\vb*{\epsilon}_\mu&=&\partial_\mu\mathcal{D}\left(\vb*{\hat{n}}_1\cos\Omega+\vb*{\hat{n}}_2\sin\Omega\right) +\mathcal{D}\,\partial_\mu\Omega \left(-\vb*{\hat{n}}_1\sin\Omega +\vb*{\hat{n}}_2\cos\Omega \right) = \partial_\mu\mathcal{D}\,\vb*{\hat{e}}_r + \mathcal{D}\partial_\mu\Omega \,\vb*{\hat{e}}_\phi,\\
\vb*{\epsilon}_\nu&=&\partial_\nu\mathcal{D}\left(\vb*{\hat{n}}_1\cos\Omega +\vb*{\hat{n}}_2\sin\Omega\right) +\mathcal{D}\,\partial_\nu\Omega \left(-\vb*{\hat{n}}_1\sin\Omega +\vb*{\hat{n}}_2\cos\Omega\right)= \partial_\nu\mathcal{D}\,\vb*{\hat{e}}_r + \mathcal{D}\partial_\nu\Omega \,\vb*{\hat{e}}_\phi,\\
\vb*{\epsilon}_s&=&\partial_s\mathcal{D}\,(\vb*{\hat{n}}_1\cos\Omega +\vb*{\hat{n}}_2\sin\Omega ) +\mathcal{D}\,\partial_s\Omega\,(-\vb*{\hat{n}}_1\sin\Omega+\vb*{\hat{n}}_2\cos\Omega ) +  v\mathcal{K}\vb*{\hat{t}}\\
&=&\partial_s\mathcal{D}\,\vb*{\hat{e}}_r+\mathcal{D}\partial_s\Omega\,\vb*{\hat{e}}_\phi+v\,\mathcal{K}\,\vb*{\hat{e}}_z,\nonumber
\end{eqnarray}
where we define the local curvature factor $\mathcal{K}(\mu,\nu,s)=1-\mathcal{D}(k_1\cos\Omega + k_2\sin\Omega)$ and also make use of the underlying local cylindrical orthonormal basis vectors $\{\vb*{\hat{e}}_r,\vb*{\hat{e}}_\phi,\vb*{\hat{e}}_z\}$ that can be constructed by aligning the axis of a cylindrical coordinate system with the tangent vector $\hat{\vb*{t}}$. This can be helpful in understanding the geometry and for basic calculations. 
From the expressions of the basis vectors we can already directly see that none of these are necessarily orthogonal. We can now also evaluate the determinant of the metric tensor $\g$ , which has following form:
\begin{eqnarray}
\g &=& \,\vb*{\epsilon}_\mu \cdot\left(\vb*{\epsilon}_\nu\cross\vb*{\epsilon}_s\right) = v\,\mathcal{D} \mathcal{K}\left(\partial_\mu\mathcal{D}\,\partial_\nu\Omega - \partial_\nu\mathcal{D}\,\partial_\mu\Omega\right).\label{eq:metric_det}
\end{eqnarray}
Since $\g$ must always be non-zero, except at the magnetic axis, we are given three separate conditions that must always be true for our coordinate system to be valid:
\begin{eqnarray}
\label{eq:cs_constraint_0}
\mathcal{D} &>& 0,\\
\label{eq:cs_constraint_1}
\mathcal{K} &>& 0,\\
\label{eq:cs_constraint_2}
\partial_\mu\mathcal{D}\partial_\nu\Omega &-& \partial_\nu\mathcal{D}\,\partial_\mu\Omega > 0.
\end{eqnarray}
The second condition limits either the size or the curvature of the flux rope, and forbids the flux rope to be bent so far that the outer boundary surface intersects with itself. The third condition forbids the intersection of different flux surfaces due to changes along the axis irregardless of the curvature. For cases where $\Omega$ is independent of $\mu$, it follows that both $\mathcal{D}$ and $\Omega$ must be strictly monotonically increasing functions for the coordinates $\mu$ and $\nu$ respectively.

It is important to understand that these constraints are limitations of our coordinate system, and it is fairly straightforward to create counter examples with an axis that is bent further, or a flux rope that is larger, than these conditions would suggest is allowed. One of the simplest examples is a flux rope that is kinked internally, so that the magnetic axis and inner flux surfaces are writhed, but the outer flux surfaces remain unperturbed. This outer flux surface could be at any distance from the magnetic axis. But according to our  constraints, and assuming a circular cross-section, the maximum radius of this flux rope is limited to $\kappa^{-1}$ which is not the case in our example. An approach one could use to tackle this issue is to ``smooth'' out $\vb*{\gamma}$ for larger $\mu$ so that $\vb*{\gamma}$ is now a function of both $\mu$ and $s$. The magnetic axis itself does not change, but it would appear to do so for the outer flux surfaces.

\subsection{Currents \& Boundary Conditions}

The magnetic field that is described by the Equation \eqref{eq:b_full} is by construction solenoidal and confined within the flux rope structure that is given by the outer flux surface at $\mu=1$. The same cannot be said for the current, which is related to the magnetic field by the following three equations:
\begin{eqnarray}
\mu_0 J^\mu =& \left(\curl \vb*{B}\right)^\mu &= g^{-\nicefrac{1}{2}} \left[\partial_\nu \left(g_{\nu s} B^\nu + g_{s s} B^s\right) - \partial_s \left(g_{\nu \nu} B^\nu + g_{\nu s}  B^s\right)\right],\label{eq:ampere_law_mu}\\
\mu_0 J^\nu =& \left(\curl \vb*{B}\right)^\nu &= g^{-\nicefrac{1}{2}} \left[\partial_s \left(g_{\mu\nu} B^\nu + g_{\mu s} B^s\right) - \partial_\mu \left(g_{\nu s} B^\nu + g_{ss}  B^s\right)\right],\label{eq:ampere_law_nu}\\
\mu_0 J^s =& \left(\curl \vb*{B}\right)^s &= g^{-\nicefrac{1}{2}} \left[\partial_\mu \left(g_{\nu\nu} B^\nu + g_{\nu s} B^s\right) - \partial_\nu \left(g_{\mu \nu} B^\nu + g_{\mu s}  B^s\right)\right],\label{eq:ampere_law_s}
\end{eqnarray}
which is Ampere's law, in our curvilinear coordinate system, with the additional  assumption that there is no displacement current. One could now ask the question if one could find an appropriate geometry and corresponding equations for the magnetic field components so that $\eval{J^\mu}_{\mu=1}=0$. This is not the case for any of the complex heliospheric flux rope models, such as \citep[e.g.][]{isavnin2016}, or the models described in N18, W22 or NC23. The full expression in Eq. \eqref{eq:ampere_law_mu} is intractable, but we can use the trivial solution where $\eval{\vb*{B}}_{\mu=1}=0$ so that any derivatives along $\nu$ or $s$ vanish. Introducing this boundary condition creates a fully shielded flux rope, with zero net axial. This constraint is stronger than the one for the commonly used shield flux rope, where only the azimuthal field component vanishes at the boundary \citep[e.g.][]{solovev2021}.

For our purposes, we will simply ignore the boundary conditions, as is commonly done when modeling ICME flux ropes. We do want to highlight, that using the above boundary condition of a vanishing magnetic field could be used to combine multiple flux ropes in a self-consistent way. Because the field vanishes at the boundary, this will also be the case for a superposition of an arbitrary number of flux ropes. This could be used to generate scenarios, where flux ropes split or merge. In terms of our functions for describing the flux rope, this only requires that the corresponding expressions converge at some point along the axis.

\subsection{Comparison With Cylindrical Models}\label{sec:comp}

We briefly discuss how to choose appropriate expressions for $\vb*{\gamma},\ \mathcal{D},\ \Omega, \, \chi,\ \xi$ and $\zeta$ so that we can reproduce the commonly used flux rope models. It should be fairly straightforward to realise that for any of these it must be that $\zeta = 0$. A cylindrical geometry requires that $\vb*{\gamma}$ is a straight line, and in the case of a torus it will be a circle with radius $\kappa^{-1}$. For a circular cross-section, the most sensible choice for the distortion function is $\mathcal{D}=\sigma\, \mu$, where $\sigma$ is the radius of the flux rope cross-section.
As we have already formulated the basis vectors of our curvilinear coordinate systems in terms of basis vectors of a local cylindrical coordinate system, we can easily transform the magnetic field components into these locally cylindrical coordinates. We can then evaluate the corresponding magnetic field components $\{B_r,\, B_\phi,\, B_z\}$ as:
\begin{eqnarray}
B_r=\hat{\vb*{e}}_r\cdot\vb*{B} &=& \frac{\chi\,\partial_\nu\mathcal{D} + \xi\,\partial_s\mathcal{D}}{\mathcal{D} \mathcal{K}\left(\partial_\mu\mathcal{D}\,\partial_\nu\Omega - \partial_\nu\mathcal{D}\,\partial_\mu\Omega\right)},\label{eq:b_cyl_r}\\
B_\phi=\hat{\vb*{e}}_\phi\cdot\vb*{B} &=& \frac{\chi\,\partial_\nu\Omega + \xi\,\partial_s\Omega}{\mathcal{K}\left(\partial_\mu\mathcal{D}\,\partial_\nu\Omega - \partial_\nu\mathcal{D}\,\partial_\mu\Omega\right)},\label{eq:b_cyl_phi}\\
B_z=\hat{\vb*{e}}_z\cdot\vb*{B} &=& \frac{v\,\xi}{\mathcal{D}\left(\partial_\mu\mathcal{D}\,\partial_\nu\Omega - \partial_\nu\mathcal{D}\,\partial_\mu\Omega\right)},\label{eq:b_cyl_s}
\end{eqnarray} 
where we now omitted the $\zeta$ function. The simplest choice for the azimuth function is $\Omega=2\pi\nu$, where we choose the angle range for $\Omega$ as $\left[0,2\pi\right]$ and it then follows that $B_\phi=\chi/\sigma$, $B_z=v\,\xi/\left(2\pi\mu \sigma^2\right)$ and $B_r=0$. As a result, we can easily recover the linear force-free flux rope model, with scalar $\alpha$, by choosing:
\begin{linenomath*}
\begin{equation}
\chi_\textrm{LFF}=\sigma B_0 J_1(\alpha \mu),\ \xi_\textrm{LFF}=\frac{2\pi \mu \sigma^2}{\eval{v}_{s=s_0}} B_0 J_0(\alpha \mu),
\end{equation}
\end{linenomath*}
or the uniform twist force-free model, with twist number $\tau$, by using:
\begin{linenomath*}
\begin{equation}\label{eq:utff}
\chi_\textrm{UT}= \frac{\mu\sigma \tau B_0}{1+\mu^2\tau^2},\ \xi_\textrm{UT}=\frac{2\pi\mu \sigma^2 B_0}{\eval{v}_{s=s_0}\left(1+\mu^2\tau^2\right)},
\end{equation}
\end{linenomath*}
where $B_0$ is the magnetic field strength at the magnetic axis. The curve velocity $v$ is a function of the $s$ coordinate, and thus must be evaluated for some fixed position $s_0$ so that the axial flux is conserved anywhere along the axis. Next, we can take a look for a torus geometry. The $B_z$ component has no dependency on $\mathcal{K}$, but any force-free torus solution must have the property that $\partial_\nu\left(g_{ss} B^s\right)=0$ which is the only term from Eq. \eqref{eq:ampere_law_mu} that does not automatically vanish. In our case, we can compute $g_{ss}=\mathcal{K}^2$, so that the previous condition transforms into $\partial_\nu\left(\mathcal{K}/\partial_\nu\Omega\right)=0$ if we let $\Omega$ be unknown. Because $\mathcal{K}$ itself depends on $\Omega$, this is a differential equation with the following solution when taking the boundary conditions into account:
\begin{linenomath*}
\begin{equation}\label{eq:omega_torus}
\Omega = 2 \arctan\left(\frac{\sigma\mu k_2 + \sqrt{1-\sigma^2\mu^2\left(k_1^2+k_2^2\right)}\tan\left(\pi\nu + \frac{\pi}{2}\right)}{1+\sigma\mu k_1}\right) + \pi.
\end{equation}
\end{linenomath*}
Note that this result is dependent on the specific choice for our normal vectors because the two curvatures values $k_1$ and $k_2$ are not interchangeable with each other. Using the previous expressions for the linear or uniform twist force free field in a cylindrical geometry, this will create a nearly force-free toroidal configuration.

The reason why we are interested in these force-free expressions is that we can use them to create more or less force-free distorted flux ropes. For the case of a distorted cross-section, we will later on show how we can circularize the cross-section geometry near the magnetic axis. Combining this circularized cross-section with a force-free solution for a toroidal geometry will yield a flux rope that is largely force-free in the center.

\section{Model Application} \label{sec:appl}

We now have all ingredients necessary to describe an arbitrarily distorted flux rope. All that is left is to choose appropriate expressions for the functions $\vb*{\gamma},\ \mathcal{D},\ \Omega, \, \chi,\ \xi$ so that we can attempt to reconstruct real \textit{in situ} events. We want to highlight here, that we expect our model to work with a variety of different shapes and magnetic field configurations, and that the demonstrated example in this manuscript is only one of the many possible realizations. 

A new aspect in this model that needs to be considered is that the overall geometry of the flux rope can change considerably depending on the orientation of the cross-section. This orientation is dependent on the particular choices for $\mathcal{D},\, \Omega$ and the normal vectors $\vb*{\hat{n}}_{1/2}$. The easiest way to control the orientation, would be by controlling the orientation of the normal vectors. But unfortunately, the normal vectors evolve along the axis according Eqs. (\ref{eq:dtds}-\ref{eq:dn2ds}) and are fixed up to a single degree of rotation. We are thus not able to control their orientation except at a single location along the axis. Instead, we will make use of the fact that there is an infinite number of different parallel transport frames, which differ by an angle of rotation. When locally evaluating any of our quantities we can always choose a particular frame so that the normal vector $\vb*{\hat{n}}_1$ points in the desired direction. Because we want to preserve some overall sense of spherical symmetry, we introduce a so-called frame anchor which we fix at the position $a=\vb*{\gamma}(0.5)/2$. In the case where our magnetic axis describes a circle, for a toroidal geometry, this is the center point. We can now always re-orientate our local parallel transport frame so that $\vb*{\hat{n}}_1$ more or less points towards $a$ by applying a simple orthogonalization procedure to the vector $a-\vb*{\gamma}$ with the added constraint that $\vb*{\hat{n}}_1$ must be normal to $\vb*{t}$. One issue with this approach is that any derivatives with respect to the $s$ coordinate are harder to calculate due to an additional rotation of the overall cross-section, and for numerical simplicity we will make the assumption that this additional discrepancy is small and we will ignore it in our calculations. For evaluation of the magnetic field this assumption is justified because our magnetic axis will not behave erratically so that the additional rotation is expected to be small. In our computation of the field lines, we included some additional ad hoc corrections so that the field lines are constrained to a particular flux surface as the errors would otherwise add up during the integration process.

For the distortion function $\mathcal{D}$, we will now use the following expression:
\begin{linenomath*}
\begin{equation}\label{eq:cross_section}
\mathcal{D}= \frac{\mu\,\sigma \delta_f \norm{\vb*{\gamma}}}{\eval{\norm{\vb*{\gamma}}}_{s=0.5}}\left\lbrace\begin{array}{ll}
     \left[\cos^2\left( \Omega + \psi \right) + \Gamma_f(\mu,s) \sin^2 \left( \Omega + \psi\right)\right]^{-\nicefrac{1}{2}},& \textrm{if } \Omega+ \psi<\frac{\pi}{2} \textrm{ or } \Omega+ \psi > \frac{3\pi}{2} \\
     \left[\cos^2\left( \Omega + \psi \right) + \Gamma_b(\mu,s) \sin^2 \left( \Omega  + \psi\right)\right]^{-\nicefrac{1}{2}},& \textrm{else}
\end{array}\right.
\end{equation}
\end{linenomath*}
that describes a semi-elliptical cross-section, where $\sigma$ is the width at $s=0.5$, when the cross-section is circular, $\Gamma_{f/b}$ is the front and back aspect ratio that varies along the axis and with respect to the distance from the axis. We also include an additional parameter $\psi$ with which we can rotate the overall cross-section. The axis and distance dependency of $\Gamma$ is chosen as:
\begin{linenomath*}
\begin{equation}
\Gamma_{f/b}(\mu,s) = 1 + \mu(\delta_{f/b} - 1)\sin^2\left(\pi\,s\right),
\end{equation}
\end{linenomath*}
where $\delta_{f/b}$ is the cross-section aspect ratio at $\mu=1$ and $s=0.5$ for both cases. and the cross-section is circularized near the magnetic axis and at the legs. The height of the cross-section at $s=0.5$ is then given by $\sigma \delta$. For the azimuth function we will simply make use of the expression in Eq. \eqref{eq:omega_torus}.

\begin{figure}[ht]
    \centering
    \includegraphics[width=0.45\linewidth]{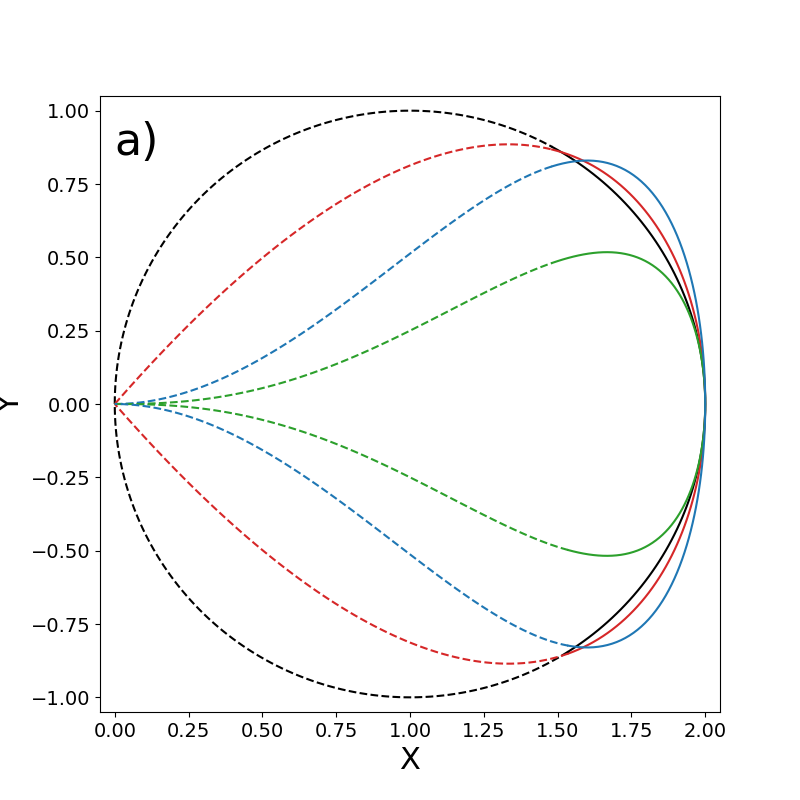}
    \includegraphics[width=0.45\linewidth]{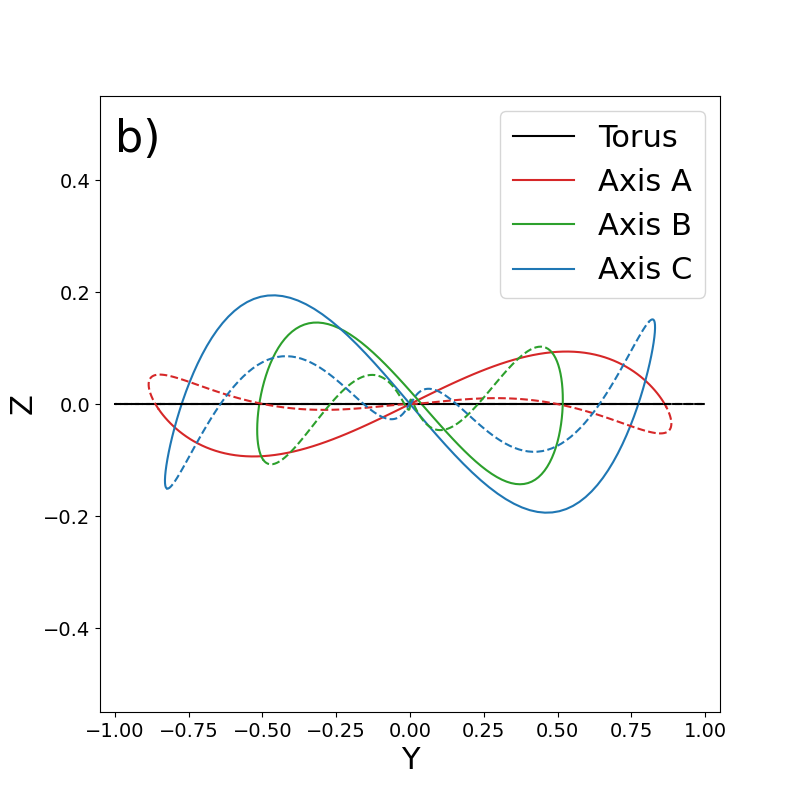}
    \caption{Examples for different parametrizations of a magnetic axis using the expression in Eq. \eqref{eq:gamma}. The solid lines show the front part of the magnetic axis with $s\in[1/3,2/3]$ and the dashed lines show the back sections. The black line shows the standard example of a torus geometry with circular magnetic axis $(\alpha=1,\beta=\lambda=\epsilon=\kappa=0)$. The red line shows an example with $(\alpha=1.15,\beta=0.1,\lambda=1,\epsilon=6,\kappa=0)$, the green line with $(\alpha=1,\beta=0.15,\lambda=4,\epsilon=8,\kappa=0.5)$ and blue with $(\alpha=1.45,\beta=0.2,\lambda=3,\epsilon=9,\kappa=1)$. The left panel is a top-down view and the right panel shows a view from the front.}
    \label{fig:gammas} 
\end{figure}

Next we describe the expression for the magnetic axis $\gamma$. In a similar way to $\mathcal{D}$ or $\Omega$, we ideally want a single expression that can capture a wide range of properties but also is similar to existing flux rope models so that one can do simple comparisons. We do not want to use composite shapes \citep[e.g.][]{janvier2013}, where we have to stitch together different curves which is hard to do with sufficient smoothness so that we do not run into issues generating our parallel transport frame. For example, the graduated cylindrical shell (GCS) model shape \citep{thernisien2006,thernisien2009} cannot be used as it is improperly stitched together and not sufficiently smooth. Instead, we will make use use a slight modification of a torus geometry, with the expression for $\gamma$ taking the following form:
\begin{linenomath*}
\begin{equation}\label{eq:gamma}
\vb*{\gamma}=\left\lbrace\begin{array}{c}
1-\cos\left(2\pi\,s\right)\\
\alpha\sin\left(2\pi\,s\right)\sin^\lambda\left(\pi\,s\right)\\
\beta \sin\left(\epsilon \pi s + \kappa\pi\right)\sin^2\left(\pi\,s\right)
\end{array}\right.,
\end{equation}
\end{linenomath*}
where $\alpha$ and $\lambda$ determine the width and front-flattening of the structure, and  $\beta,\,\epsilon,\, \kappa$ are three parameters that determine the variation of the axis in the third dimension (z-axis w.r.t to the ICME plane). This shape is not necessarily extremely realistic, but is simple to describe and behaves well when implemented numerically. With the above example, we receive a circle by choosing  $\alpha=1$ and $\beta=\lambda=0$ which can be used to describe a torus. While our model in principle can handle more complex shapes, with much more sophisticated parametrizations, it turns out that handling shapes with highly varying curve velocities $v$ can be numerically problematic. An example where this occurs is the Fri3D axis from \cite{isavnin2016}, where $v$ tends to infinity at both ends of the flux rope. Figure \ref{fig:gammas} shows four different examples of a magnetic axis as it can be described using Eq. \eqref{eq:gamma}. The first example is the standard torus geometry, that has a circular magnetic axis. The other three axes use a variety of different model parameters that are explained in detail in the caption. In Fig. \ref{fig:gammas}b we use different line styles to denote the front (solid) and back (dashed) portion of the axis due to the lack of depth perception. From these figures we can already extract a few key properties of our model, namely regardless of the parameters the apex point on the axis will be the same. The same holds for the curvature of the flux rope at this same apex point. 

\begin{figure}[ht]
    \centering
    \includegraphics[width=0.95\linewidth]{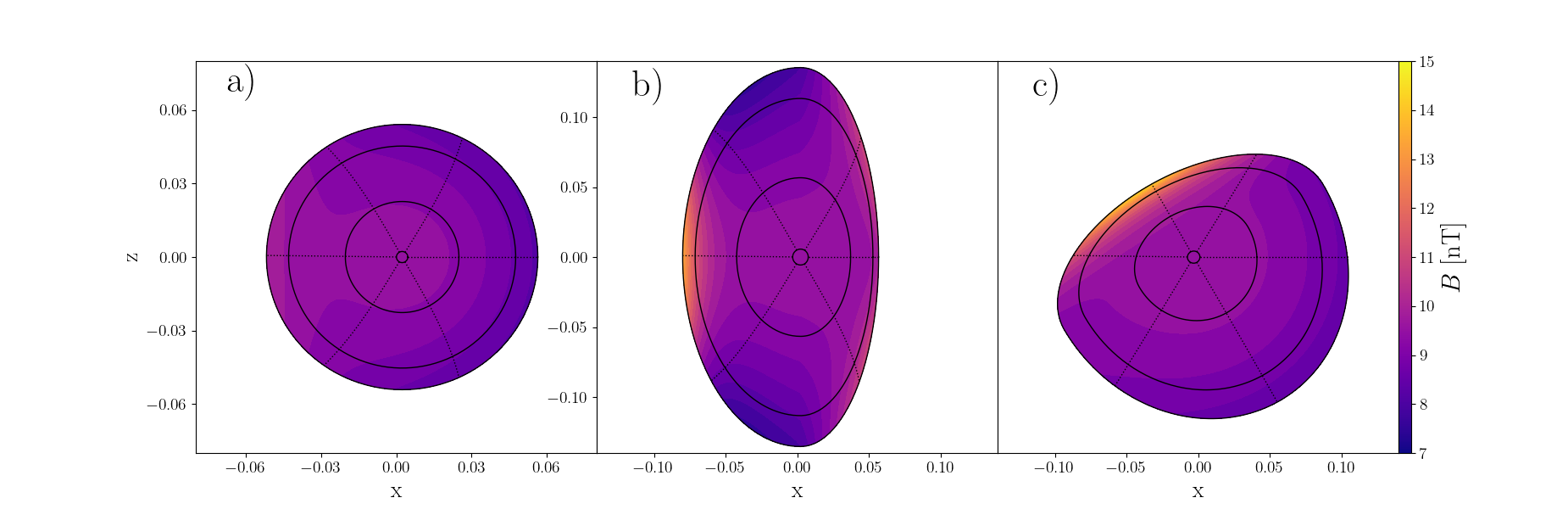}
    \includegraphics[width=0.95\linewidth]{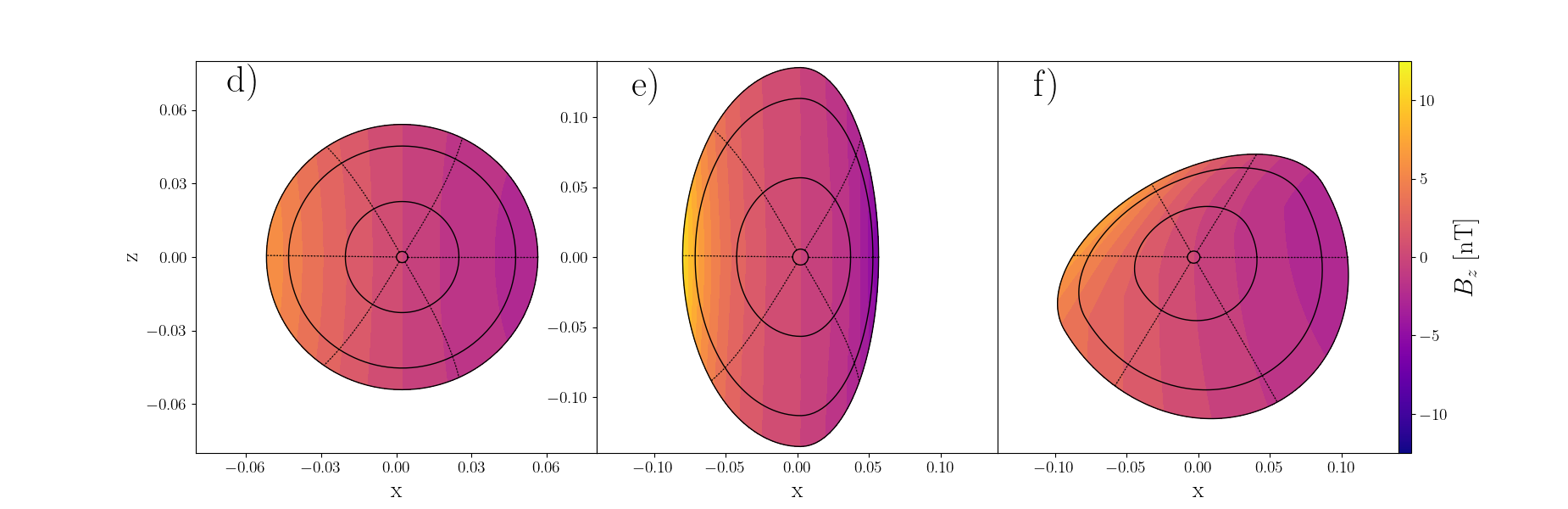}
    \caption{Cross-sections for three different examples with different aspect ratios. Panels (a-c) show the total magnetic field strength $|B|$. Panels (d-f) show the magnetic field strength along the $z$-axis. Panel a/d shows a standard circular cross-section which is the same as any Gold-Hoyle model. Panel b/e shows a cross-section with $\delta_f=2.5$ and $\delta_b=1.66$. The positive X-axis points forward, so that the compressed part of each cross-section is on the left side of each panel. Panel c/d shows a cross-section with $\delta_f=1.75$ and $\delta_b=0.875$ and $\psi=2\pi/3$}.
    \label{fig:cross_sections}
\end{figure}

Lastly, we need to choose expressions for $\chi$ and $\xi$, where we want to largely use the existing expressions from \eqref{eq:utff} with minor modifications. As the resulting magnetic field in the end also depends on the geometry, it can be useful to also include these dependencies in the two functions $\chi$ and $\xi$ directly in the hope that the geometry terms balance out and reproduce a more similar field no matter which parameters are chosen. As such, we will use the following expressions for the magnetic field functions:
\begin{linenomath*}
\begin{equation}\label{eq:utffmod}
\chi = B_0\frac{\eval{\mathcal{D}}_{\nu=0} \tau}{1+\mu^2\tau^2},\ 
\xi = B_0\frac{ 2\pi\eval{\mathcal{D}}_{s=s_0}\eval{\partial_\mu\mathcal{D}}_{s=0.5}}{1+\mu^2\tau^2},
\end{equation}
\end{linenomath*}
where we are using the force-free uniform twist field as an inspiration. We replaced some of the factors in terms of derivatives of $\mathcal{D}$ evaluated at specific locations so that the coordinate dependencies are correct.

Figure \ref{fig:cross_sections} illustrates three different examples of a flux rope cross-section as implemented using our approach. The first panel (a/f), shows a standard circular-cross section and the second and third panels (b/c/e/f) show a double-elliptical cross-section. The first row (a-c) shows the magnetic field strength and the second row (d-f) only the $B_z$ component as a colored contour plot. The constant $\mu$ contours are shown with solid, and the constant $\nu$ contours in dashed black lines. The $\nu$ contours are slightly bent backwards as we included some curvature in these examples, where the flux ropes are bent in the negative $x$ direction. This leads to an enhancement in the magnetic field strength at the back. One can also see an enhancement in the frontal regions for the distorted cross-sections where the flux surfaces converge, and for the case shown in panel (c), the combination curvature and compression of flux surfaces creates the strongest field at $\nu=0$ which is found at the top left part of the panel.

\subsection{Integration within 3DCORE}

We implemented the above flux rope model, with our specifically chosen expressions for the geometry and magnetic field, within the existing 3DCORE framework, that is described in \cite{weiss2021a,weiss2021b}. The primary motivation for doing so is that we can use the existing Monte-Carlo fitting procedure with only minimal adaptions. As the 3DCORE approach assumes that our flux rope structure evolves in time, we do need a basic procedure to evolve our parameters, which we achieve in a similar way as for the original model by using empirical scaling relations:
\begin{eqnarray}
\gamma(t) &=& \rho_0(t)\gamma,\\
\mathcal{D}(t) &=& \rho_1(t) \mathcal{D},\\
B_0(t) &=& B_0 \left(2\rho_0\right)^{-n_b},\\
\tau(t) &=& \frac{\tau}{\int\dd{s}v},
\end{eqnarray}
where $\rho_0$ and $\rho_1$ are the major and minor radii of the underlying torus that behave as in the original 3DCORE setting. The magnetic decay rate $n_b$ is in most cases fixed to $1.64$. We make use of the same simplified drag model and an isotropic solar wind background as in the original papers. Using the settings $\alpha=1$ and $\beta=\lambda=0$ for $\gamma$ we can recover the exact geometry as in 3DCORE which is useful for direct comparisons. 

In order to evaluate the magnetic field in a given heliospheric coordinate system, we are required to implement coordinate transformation functions that convert Cartesian coordinates into our distorted coordinates. This is done using a two step procedure. First we determine the appropriate $s$ coordinate by finding the closest position on $\vb*{\gamma}$ to the given point in space using a combined bisection and Newton method approach. Given a good estimate for $s$, we solve for both $(\mu,\,\nu)$ simultaneously using a damped Newton method from multiple starting positions to increase the chance of finding a proper solution. In the current implementation that we used to generate the results in this manuscript, this overall procedure is around one to two magnitudes slower than in the two first 3DCORE papers. Given the fact that the fitting procedure in W21b took only a few minutes for single-point events, we can achieve similar results within 10 to 30 minutes on a typical machine if we sample a limited parameter space. Given proper optimizations, we do expect the run times to improve significantly in the future.

The expression for $\vb*{\gamma}$, assumes that the ICME propagates in the direction given by the positive X-axis. Similarly to the way the tapered torus was handled in the original 3DCORE model, we thus require three additional parameters to define the propagation direction and inclination. In total, the resulting model as it is implemented in our code has effectively 20 model parameters. We will list them here for completeness sake, but it should be noted that for any reconstruction at least half of these will be fixed to some given value that can be inferred from either observations or an empirical study. The first three parameters concern the propagation direction and orientation, namely the longitude $lon$, latitude $lat$ and inclination $inc$. Note that due to the added complexity within $\vb*{\gamma}$ with different values for $\epsilon$ or $\kappa$, the local inclination of the magnetic axis can vary substantially from $inc$. The next parameter is the flux rope width at 1~au given by $\sigma_\text{1au}$. Again it is important to remember here that this parameter only properly corresponds to its description if the cross-section is circular due to the way the distortion function is defined. The two parameters $\delta_{f/b}$ give the aspect rations of the front and back half of the cross-section, $\delta_b$ is implemented in the code as a ratio w.r.t. $\delta_f$. We also have the parameter $\psi$ that rotates the cross-section in the desired angle. The magnetic field strength at the magnetic axis at 1~au is given by $B_0$ and the twist factor $\tau$ is implemented similarly to W21b, except that the total twists over the entire structure can only be evaluated numerically. The parameters describing the drag-based propagation are the same as in W21a, namely the initial velocity $v_0$, the initial distance from the Sun $R_0$, the isotropic solar wind speed $v_\text{sw}$ and the drag coefficient $\Gamma$ (not be confused with the effective cross-sections aspect-ratio). The scaling relation for the magnetic field $n_b$ is typically set to $1.64$ and the scaling relation for the width, given by $n_a$, is set to $1.14$. Lastly we have the five new parameters that determine $\vb*{\gamma}$, which are $\alpha$, $\beta$, $\lambda$, $\epsilon$ and $\kappa$.


\begin{figure}[ht]
	\centering
    \includegraphics[width=0.46\linewidth,trim={250 150 250 50},clip]{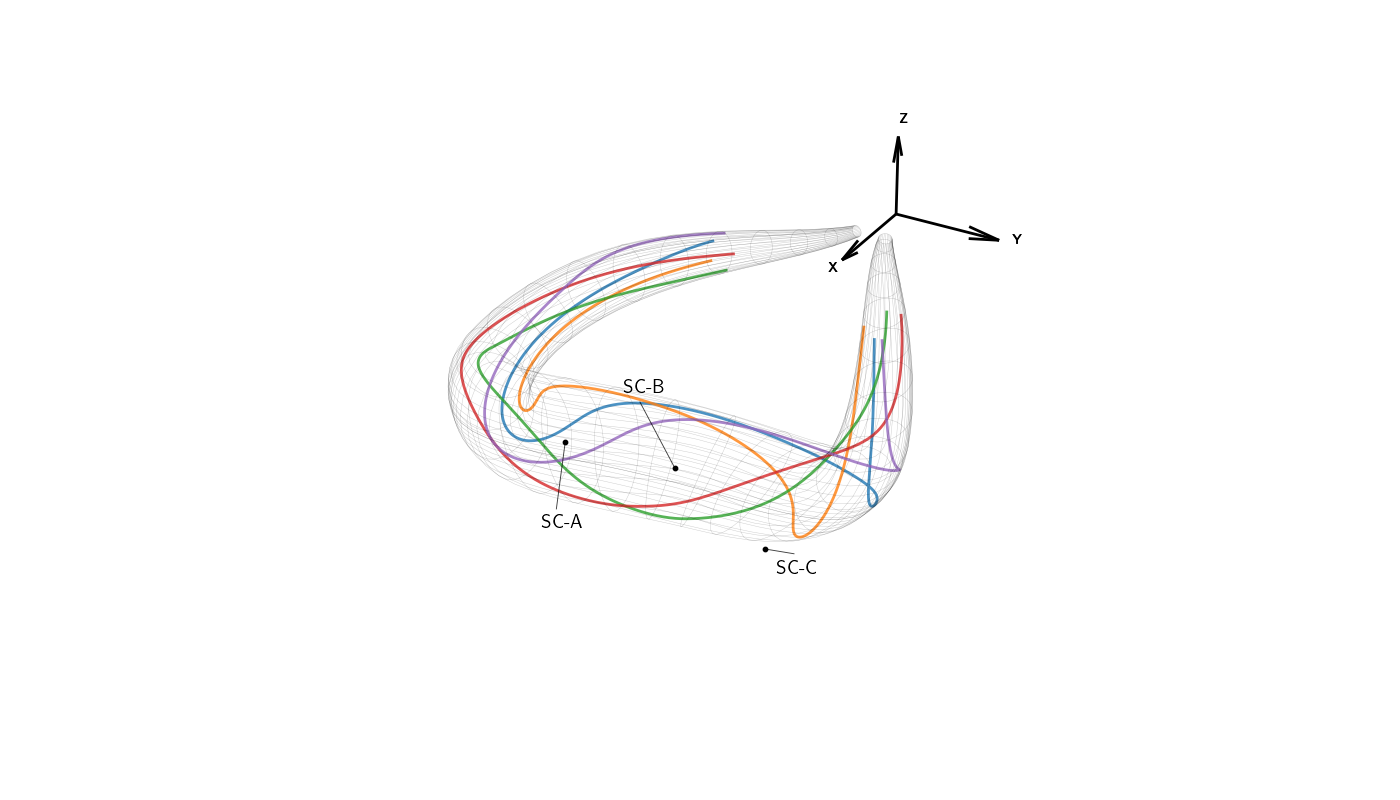}
    \includegraphics[width=0.45\linewidth,trim={250 150 225 100},clip]{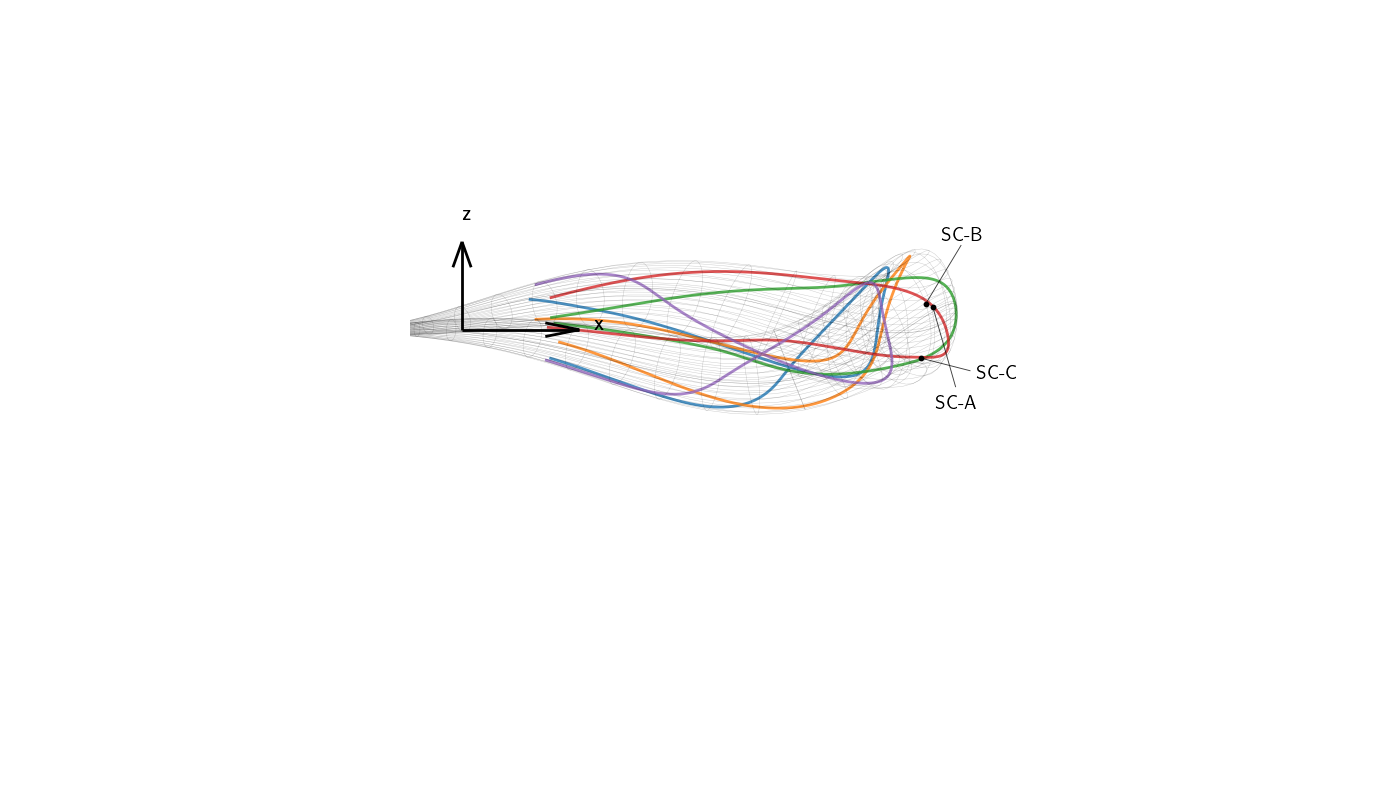}
    \includegraphics[width=0.5\linewidth,trim={25 0 50 50},clip]{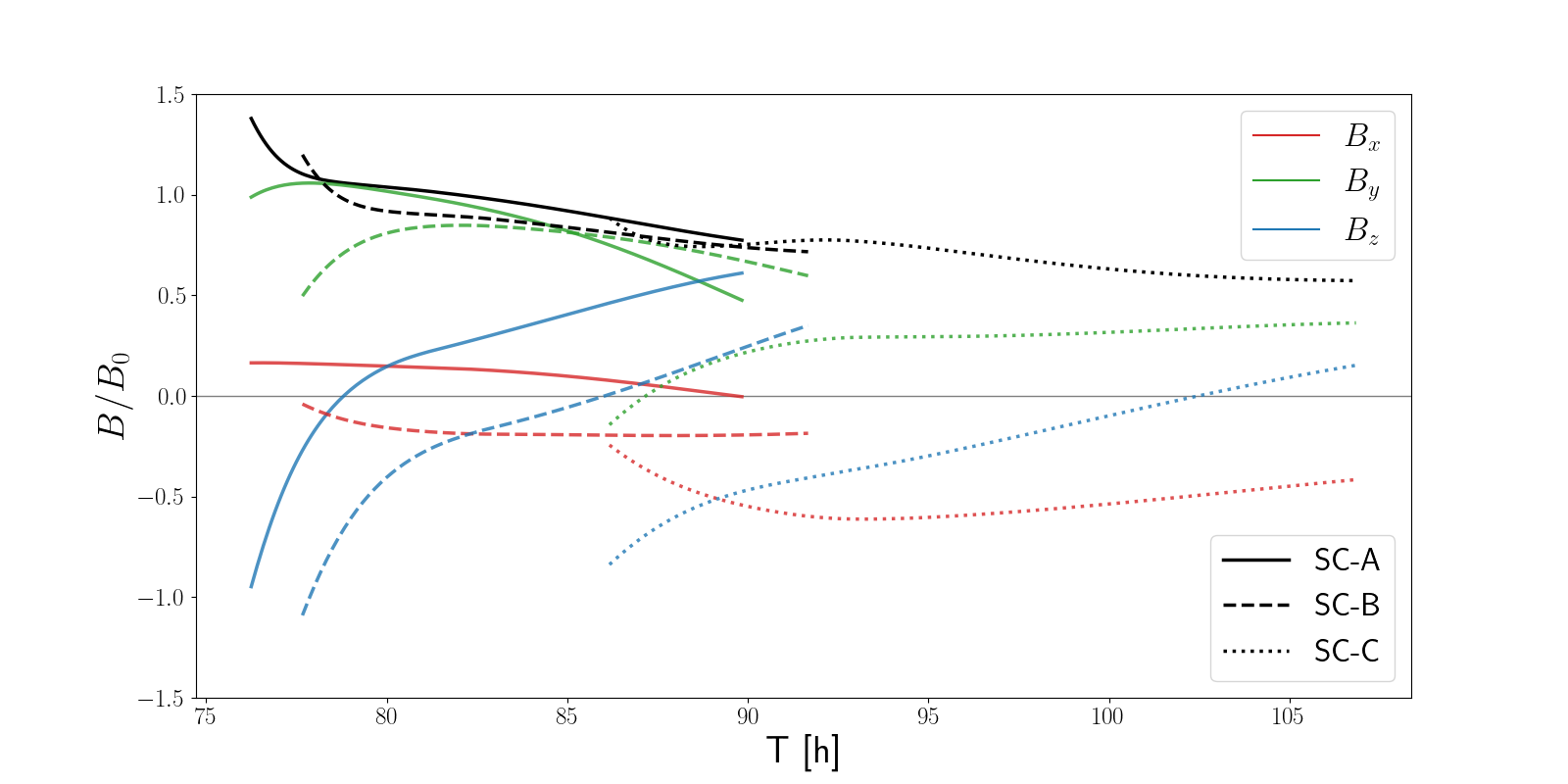}
    \caption{Exemplary implementation of our model with given expressions within the overall 3DCORE framework. \textit{Top panels:} Three-dimensional geometry of the overall flux rope structure with a few (colored) field lines and three imaginary spacecraft (SC-A/B/C). The black arrows define a Cartesian coordinate system. \textit{Bottom panel:} Three different synthetic in-situ magnetic field profiles generated at the fixed locations denoted in the top panels, using the same Cartesian coordinate system.}
    \label{fig:3dcore_impl}
\end{figure}

Figure \ref{fig:3dcore_impl} shows an example of how an ICME flux rope will look using our 3DCORE implementation and three different virtual spacecraft. For this particular example we have used $\delta_f=2.5,\ \delta_b=0.66,\ \alpha=1,\ \beta=0.15, \ \lambda=2,\ \epsilon=6\,\ \kappa=0.25$ and $\psi=-0.05$. The two top panels show the overall geometry of our flux rope at approximately 80 hours after initialization at 8 solar radii, including a number of different field lines at $\mu=0.9$ from $s=0.25$ to $=0.75$. We inserted three static virtual spacecraft in front of the ICME shape to generate synthetic \textit{in situ} measurements that are shown in the right panel of the same figure. The magnetic field profile is shown in the same coordinate Cartesian coordinate system that is defined in the left and center panels.

In these \textit{in situ} measurements, we can clearly see a compression at the front, and an asymmetric rotation of the magnetic field components. The SC-C observer has been specifically positioned to better capture the scenario of a flank encounter with a strong $B_x$ component.

\subsection{Event Analysis}

In order to demonstrate the usefulness of our overall model to real events we will apply the DMFR to a recent multi-point ICME observation. We have selected a particular ICME event that was observed by the Wind and STEREO-A spacecraft, at a suitable separation distance of around 10 degrees in heliospheric longitude in order to see differences in their magnetic field profiles \citep[e.g.][]{lugaz2024}. This CME was observed remotely by SOHO/LASCO/C3 onward from 2023 April 21 18:12 UTC (LASCO CME catalog\footnote{\url{https://cdaw.gsfc.nasa.gov/movie/make_javamovie.php?stime=20230421_1655&etime=20230421_2049&img1=lasc2rdf&title=20230421.181206.p180g;V=1284km/s}}). Due to the nature of halo CMEs, it is hard to draw any conclusions about the propagation direction or the orientation of the overall structure. The initial speed, within the cited catalog, is given as 1216~km~s$^{-1}$ at 20 $R_s$, thus it was a fast CME. After the shock arrival time at STEREO-A at 2023 Apr 23 14:29 UT, STEREO-A was positioned at 0.964 au and -10.2 degrees heliospheric longitude (HEEQ) and -5.8 heliospheric latitude when it entered a magnetic obstacle on Apr 23 20:30 UT, as given in the HELIO4CAST ICMECAT\footnote{\url{https://helioforecast.space/icmecat}} 
catalog \citep{moestl2020}. At Wind, the shock arrived 2 hours and 33 minutes later on 23 Apr 23 17:02 UT, with Wind at 0.997 au and at -0.1 degree longitude (HEEQ) and -4.92 degree latitude when it entered a magnetic obstacle on Apr 24 01:00 UT. In summary, STEREO-A was situated about 10 degrees east of the Sun--Earth line and only at about 1 degree difference in latitude. 

\begin{figure}[ht]
    \centering
    \includegraphics[width=\linewidth,trim={100 25 75 100},clip]{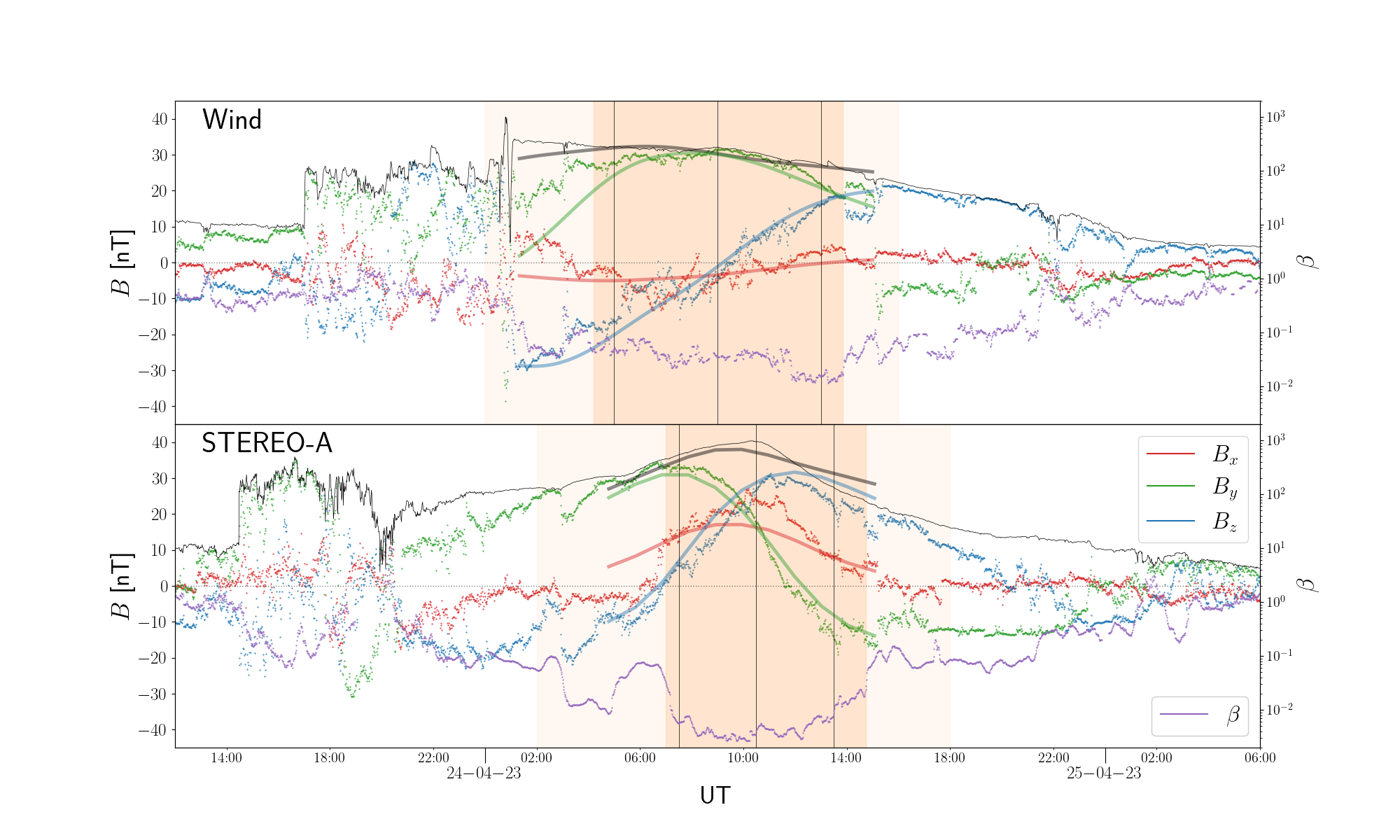}
    \caption{\textit{In situ} spacecraft measurements of an ICME passing over Wind (top panel) and STEREO-A between the 23rd to 25th of April 2023. For both spacecraft, we show in the magnetic field in HEEQ coordinates with the magnetic field components $B_x$ (red), $B_y$ (green), $B_z$ (blue) and the plasma beta parameter $\beta$ (purple). The dark orange region represents the magnetic cloud interval that we identify based on the magnetic field behaviour and the $\beta$ parameter. For Wind the magnetic cloud period was identified as 2024 April 24 04:12 to 13:51 UT. For STEREO-A it is 2024 April 24 06:59 to 14:44 UT. The light orange region is the interval that we allow our Monte-Carlo algorithm to produce outputs for, which are then compared at only three different locations which are marked by a solid horizontal line. From the results, the three solid lines in each panel, we see that this is sufficient to generate a highly accurate reconstruction of the magnetic field using the DMFR. Note that these two reconstructions were performed completely independently. }
    \label{fig:data}
\end{figure}

Figure \ref{fig:data} shows the \textit{in situ} observations from both the Wind and STEREO-A spacecraft and a specifically chosen single-point reconstruction using our DMFR with the sequential Monte-Carlo algorithm. The magnetic field measurements (colored in red, green and blue) are shown in the HEEQ coordinate system and are provided by Wind/MFI \citep{lepping1995} and STEREO-A/IMPACT \citep{acuna2008}. We also show the plasma beta parameter $\beta$ in purple, which is calculated from bulk plasma parameters, which are provided by the instruments Wind/SWE \citep{ogilvie1995} and STEREO-A/PLASTIC \citep{galvin2008}. From this, we derive our own magnetic cloud intervals within the wider magnetic obstacle at Wind from 2024 April 24 04:12 to 13:51 UT , STEREO-A from 2024 April 24 06:59 to 14:44 UT.
In both observations, we see clear indications of a well behaved ICME with a shock and sheath and a well-behaved magnetic cloud. Both the close timing of the observations and the fact that no other Earth-directed CME could be responsible for these \textit{in situ} observations establishes that the same ICME is indeed measured at both spacecraft. The peak speed of this ICME is measured as 745~km~s$^{-1}$ by Wind and 699~km~s$^{-1}$ by STEREO-A. In both cases the preceding solar wind speed was around 350~km~s$^{-1}$and the MC portions had a mean speed of 547~km~s$^{-1}$ for Wind and 551~km~s$^{-1}$ for STEREO-A. As STEREO-A was at $0.96\,au$ from the sun, the shock arrives $2.5$ hours earlier there compared to Wind, which corresponds to a mean shock speed of 519~km~s$^{-1}$.

  
But we have identified that the magnetic cloud part is seen earlier by Wind by a time difference of nearly six hours. There also appears to be separate structure in front of the MC at STEREO-A with no equivalent within the Wind measurements. We use both the magnetic field and plasma measurements to identify the time period of the magnetic cloud passage, which we have shaded in light orange. For the case of the Wind observation, we have a typical SWN flux rope with almost no radial field component. For STEREO-A, both $B_y$ and $B_z$ flip sign and there is a strong $B_x$ contribution. It can be thus expected that the geometry of the flux rope will differ significantly at both locations, which is to be expected as Wind and STEREO-A are around $0.17\,au$ apart during this time.

For both events, our reconstructions were done separately for each spacecraft (single-point).  As we do not expect many of the newly introduced model parameters to be required to reconstruct single-point observations, we fixed $\beta=\epsilon=\kappa=0$. We also limited the aspect-ratio of the cross-section to a maximum of $2$. The background solar wind speed is fixed to 350~km~s$^{-1}$ and the initial distance of the apex point is set to 20~$R_s$. In both cases, the reconstructions worked very well as can be seen from visual inspection of the \textbf{in situ} recontructions in Figure \ref{fig:data}. An interesting, but not necessarily unexpected result, is that the inclination of the magnetic flux rope is significantly different in both cases. For Wind, the reconstructed inclination was around $232^\circ$ and for STEREO-A $158^\circ$ with few degrees of variation over the ensemble. This is a $75^\circ$ difference from one spacecraft to the next. Both reconstructions also prefer slightly asymmetric cross-section rations, where $\Gamma_b/\Gamma_f \approx 0.85-0.9$.

\begin{figure}[ht]
    \centering
    \includegraphics[width=0.9\linewidth,trim={100 25 75 100},clip]{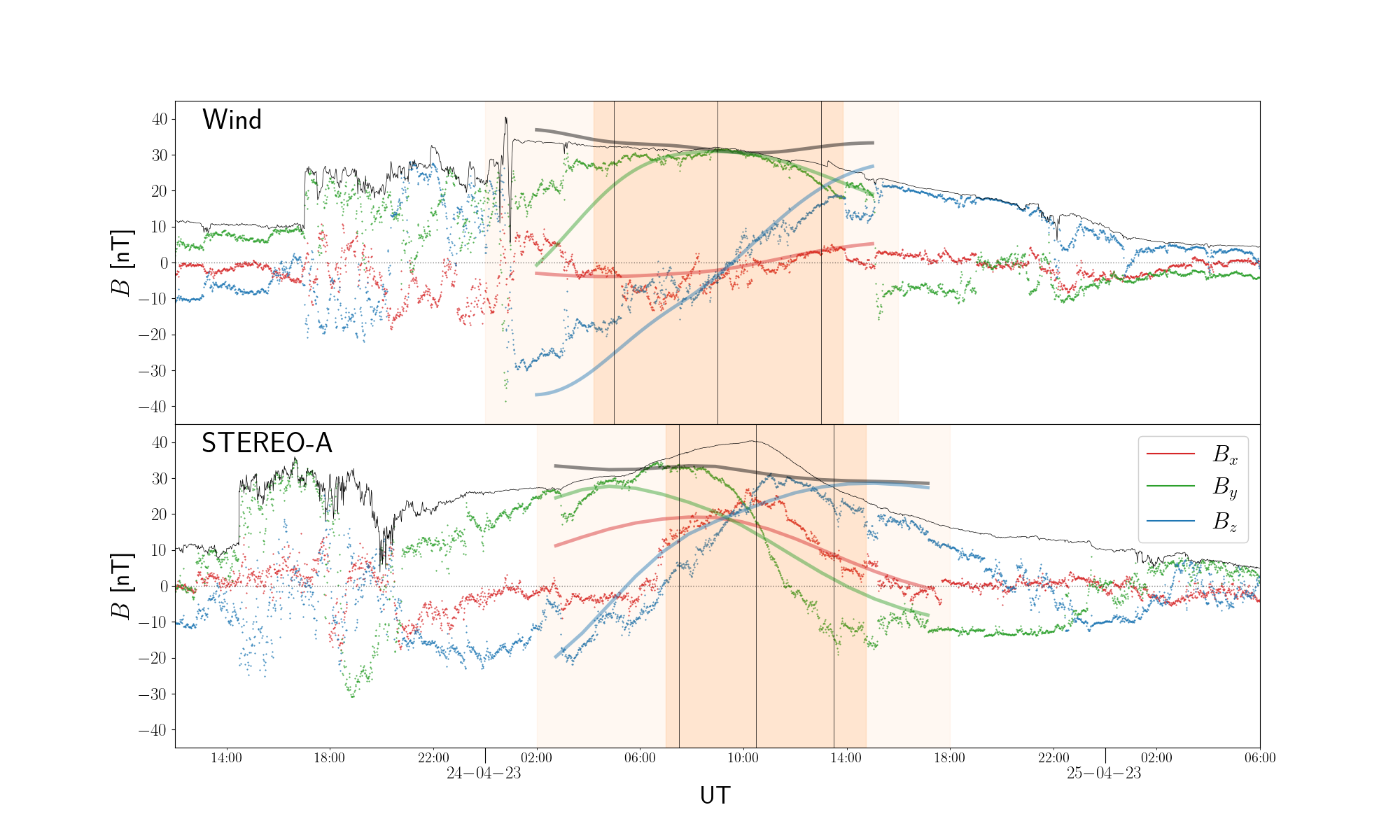}
    \includegraphics[width=0.3\linewidth,trim={325 125 300 175},clip]{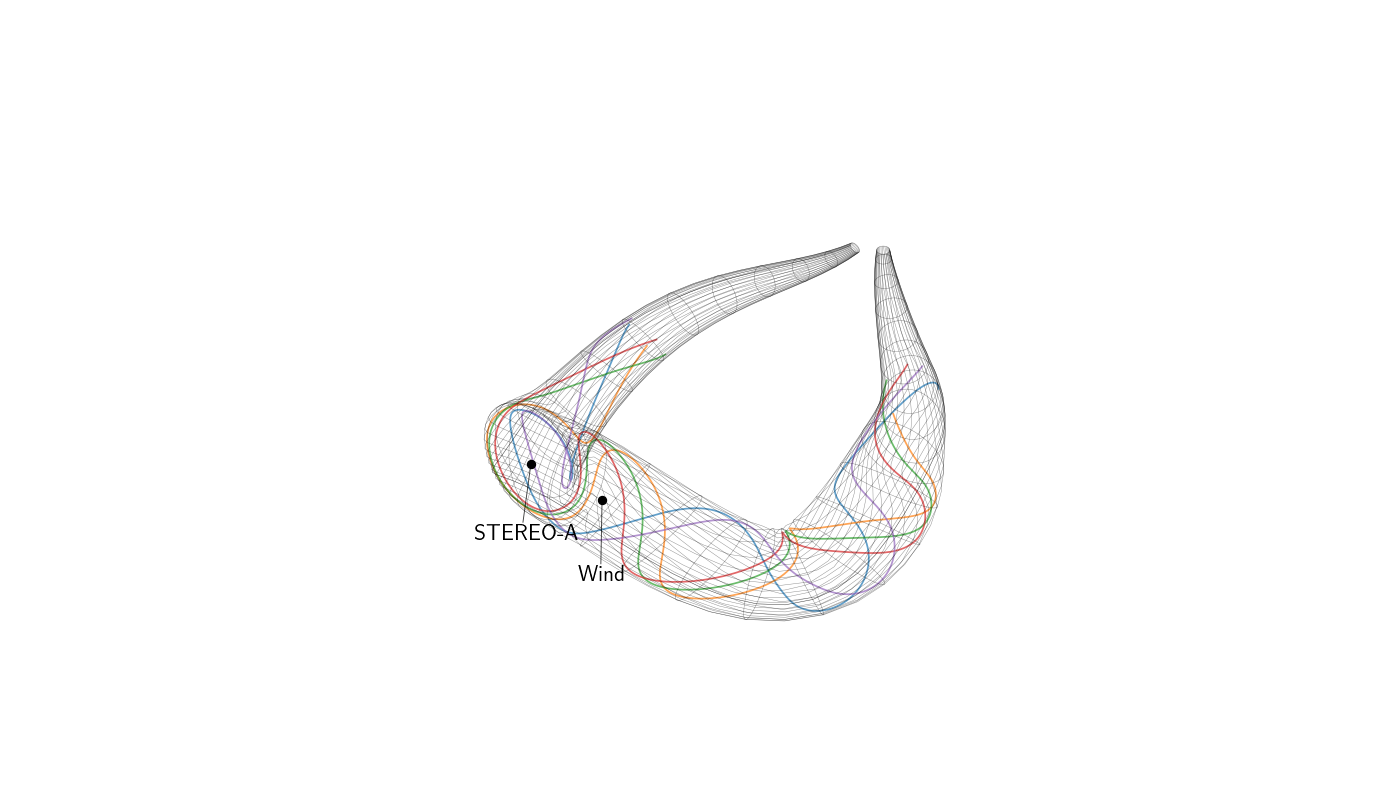}
    \includegraphics[width=0.3\linewidth,trim={325 200 275 125},clip]{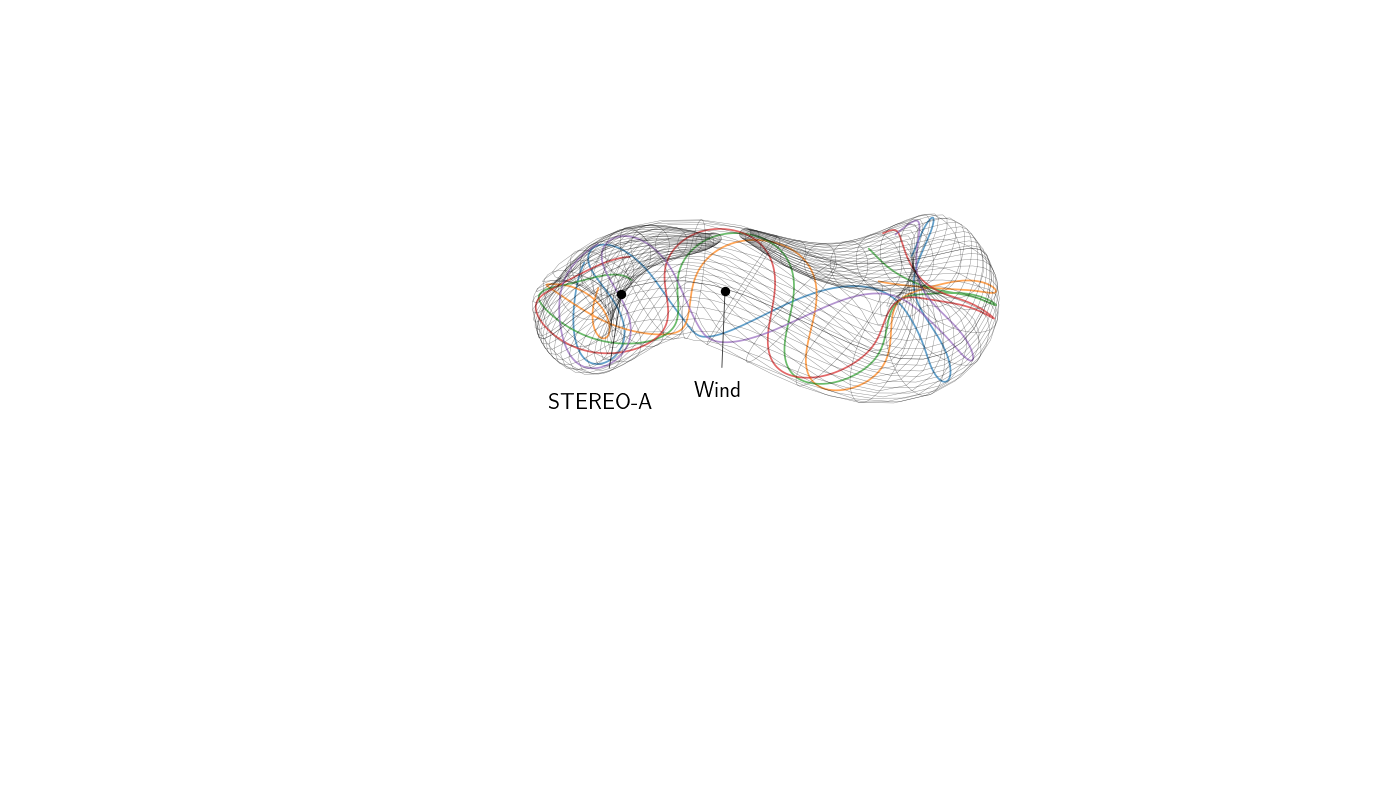}
    \includegraphics[width=0.3\linewidth,trim={325 100 275 175},clip]{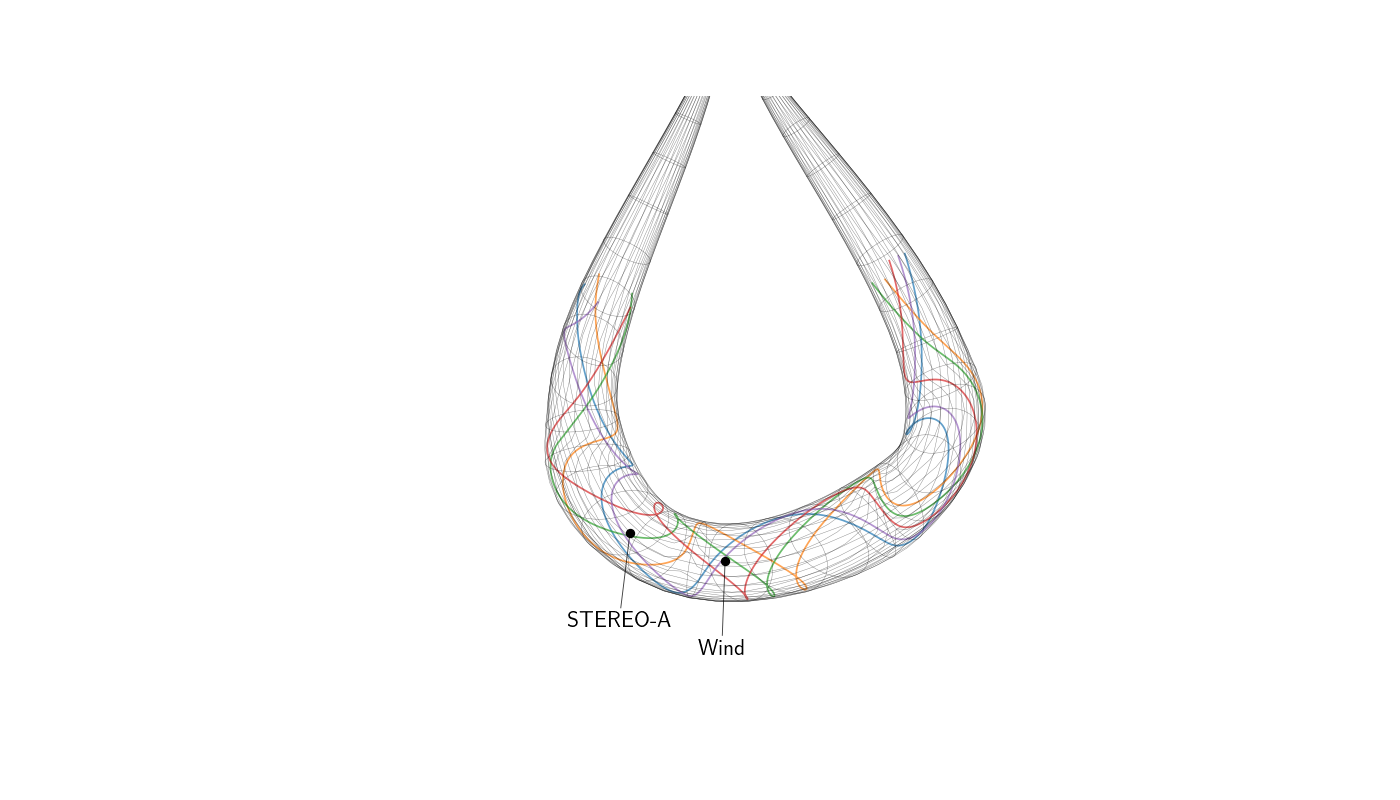}
    \caption{Results from our multi-point reconstruction using the observations from Wind and STEREO-A simultaneously. The top shows the \textit{in situ} reconstruction for both spacecraft. Here wen can visually conclude that the reconstruction works fairly well for the Wind observations, but not as good for STEREO-A when compared to the single-point reconstructions from Fig. \ref{fig:data}. We can also clearly see that the model has issues with respects to the timing in the STEREO-A data, as it expects the MFR interval to arrive earlier and it also expects the MFR interval to be much longer in duration. The bottom panels  show the global 3D shape as inferred from our model, from the side/front/top respectively, including the two spacecraft positions and a handful of illustrative field lines. Here can nicely see that the local axis inclination differs significantly in between the two spacecraft positions. In this particular example, we can calculate this angle to be $49$ degrees.}
    \label{fig:multi}
\end{figure}

Figure \ref{fig:multi} now shows our attempt and creating a multi-point reconstruction, using the same overall approach as in W21 where this was done with multiple spacecraft at much smaller distances. Because of the additional complexity of our model, with numerous additional parameters, it is not guaranteed that the fitting method will succeed without some additional manual configuration. Compared to the previous reconstructions, we now allow for a non-zero $\alpha$ parameter, but we fix $\lambda=2$, $\epsilon=6$, and $\delta_f=2$ so that the reconstruction algorithm is not overwhelmed. As can be seen for the results, we are fortunately still able to capture the overall behaviour of the observations for the Wind spacecraft, with only minor differences compared to the single-point reconstruction. The STEREO-A reconstruction is more problematic, as the model is not able to handle the timing discrepancies for our selected magnetic cloud interleaves, but still matches more or less well qualitatively. While the shock arrives earlier at STEREO-A as expected, the MFR portion that we identify arrives later than at Wind. 

The interpretation of these results is not completely clear, due to the lack of an extremely good reconstruction for the STEREO-A data. All ensemble members from our reconstruction algorithm favor the idea of a ``local'' inclination for the MFR at both spacecraft. For the example (best-fit) shown in Fig \ref{fig:multi} the difference in inclination is $49^\circ$. The lowest and largest inclination differences within our generated ensemble was $30^\circ$ and $70^\circ$ with a mean of $44^\circ$. This range is below the value of $75^\circ$ that was found in the single-point reconstructions, but nonetheless can be taken as a strong indication that the flux rope, and the associated ICME, is likely to be globally distorted into an inverse V-shape.

\section{Conclusion \& Discussion} \label{sec:conclusion}

In this manuscript we have introduced a more general framework that allows the description of almost arbitrarily complex magnetic flux rope geometries for the purpose of describing the magnetic configuration o  ICMEs. To our knowledge, this is the first flux rope model that allows for the choice of a non-circular cross-section shape and a distorted magnetic axis. With this tool in hand, it will now be possible to more properly test a number of different assumptions that various authors have made over the years concerning the local or global geometry of ICME flux rope structures or their internal magnetic field structure. For example, this model should be much more efficient at characterizing our or flank encounters as it is able to describe the transition from the front of the CME to the legs with many more degrees of freedom without having to make the assumption of constant curvature as in a toroidal model. We could also test how well our approach can differentiate between twist and writhe \citep[e.g.][]{alhaddad2019}. Some of these questions or problems will be the focus of future investigations.

While the approach itself can describe extremely complex geometries, we unfortunately believe that the only effective implementation can be achieved by parameterizing the geometry in terms of a handful of arbitrary quantities. Ideally the number of these should be limited, and there should not be too strong parameter degeneracies in b although avoiding these completely is likely to be impossible. While the example that we show in this paper, from Eq. \eqref{eq:gamma} and \eqref{eq:cross_section}, can be considered a significant improvement over existing models, they were primarily chosen for their simplicity to understand, implement and compute and may not be a very realistic representation of the structure of these ICMEs. Given further optimizations of the currently developed computer codes, it should be possible to use more sophisticated shapes that are more realistic.

With the current implementation, the overall framework is already sufficiently fast to be used for large ensemble simulations and potentially real-time space-weather forecasting applications. In the case of forecasting applications, one prohibiting factor is the increased need of initial conditions to describe and confine the parameters that describe the global shape geometry. As it is currently currently very hard to gather sufficient, or sometimes any, information on early state CME/ICME properties using either remote sensing observations or \textit{in situ} observations sub L1, it is likely not warranted to actually use this type of model until more reliable observations become available.

As a demonstration, we have shown that such a parametrized shape can be used to perform a in situ reconstruction of an ICME event at single or multiple locations. While the overall result for the multi-point reconstruction is far from perfect, it strongly suggests that the local geometry of an ICME flux rope can substantially vary at different spacecraft locations. A similar occurrence was already likely found in \cite{pal2023}, although the scenario was not as simple as for our event. With only two recent examples at hand, it is unclear how often such a strong distortion of an ICME occurs. There are a few possible explanations of how this can occur. One of these is a partial deflection of a CME, where one side of the CME is is deflected differently than the other resulting in a bent or twisted overall structure. Another possibility, is that the flux rope is kinked during the eruption phase.

The fact that the same ICME has, or can appear to have, significantly differing orientations at two different locations in space has significant implications for both space weather predictions and the overall study of ICMEs. As a conclusion, this would mean that the usage of popular cylindrical/toroidal flux rope models for analysis of multi-point ICME observations with larger spatial separations is fundamentally insufficient to capture the large-scale geometry or structure of the ICME. Because the existing fleet of spacecraft is primarily located within the Earth's ecliptic plane, this can lead to certain biases in analysis. For space weather forecasting, the existence of these large-scale distortions has the consequence that it is likely much harder to forecasting the important $B_z$. Assuming that one had a very good idea of the large-scale structure of an ICME, a small error in the estimation of the propagation direction could cause significant changes in the $B_z$ profile if the inclination axis changes sufficiently. This issue is very relevant with the respect to the estimation of basic CME properties from white-light observations \citep[e.g.][]{verbeke2023}{}. As the occurrence of multi-point observations will only increase in the future, thanks to the multitude of existing and new planned spacecraft missions, the unknown complexity of the large-scale ICME structure will have to be ever more taken into account to both improve our understanding of ICMEs and the capability of space weather forecasting.

\section*{acknowledgments}
A.J.W. acknowledges the financial support by an
appointment to the NASA Postdoctoral Program at NASA Goddard Space Flight Center, administered by Oak
Ridge Associated Universities through a contract with NASA. T.N-C. thanks for the support of the Solar Orbiter and Parker Solar Probe missions, Heliophysics Guest Investigator Grant 80NSSC23K0447 and the GSFC-Heliophysics Innovation Funds. Funded by the European Union (ERC, HELIO4CAST, 101042188). Views and opinions expressed are however those of the author(s) only and do not necessarily reflect those of the European Union or the European Research Council Executive Agency. Neither the European Union nor the granting authority can be held responsible for them.

\bibliography{refs}{}

\begin{thebibliography}{}
\expandafter\ifx\csname natexlab\endcsname\relax\def\natexlab#1{#1}\fi
\providecommand{\url}[1]{\href{#1}{#1}}
\providecommand{\dodoi}[1]{doi:~\href{http://doi.org/#1}{\nolinkurl{#1}}}
\providecommand{\doeprint}[1]{\href{http://ascl.net/#1}{\nolinkurl{http://ascl.net/#1}}}
\providecommand{\doarXiv}[1]{\href{https://arxiv.org/abs/#1}{\nolinkurl{https://arxiv.org/abs/#1}}}

\bibitem[{{Acu{\~n}a} {et~al.}(2008){Acu{\~n}a}, {Curtis}, {Scheifele}, {Russell}, {Schroeder}, {Szabo}, \& {Luhmann}}]{acuna2008}
{Acu{\~n}a}, M.~H., {Curtis}, D., {Scheifele}, J.~L., {et~al.} 2008, \ssr, 136, 203, \dodoi{10.1007/s11214-007-9259-2}

\bibitem[{{Al-Haddad} {et~al.}(2019){Al-Haddad}, {Poedts}, {Roussev}, {Farrugia}, {Yu}, \& {Lugaz}}]{alhaddad2019}
{Al-Haddad}, N., {Poedts}, S., {Roussev}, I., {et~al.} 2019, \apj, 870, 100, \dodoi{10.3847/1538-4357/aaf38d}

\bibitem[{Bishop(1975)}]{bishop1975}
Bishop, R.~L. 1975, The American Mathematical Monthly, 82, 246.
\newblock \url{http://www.jstor.org/stable/2319846}

\bibitem[{Boozer(1981)}]{boozer1981}
Boozer, A.~H. 1981, The Physics of Fluids, 24, 1999, \dodoi{10.1063/1.863297}

\bibitem[{{Bothmer} \& {Schwenn}(1998)}]{bothmer1998}
{Bothmer}, V., \& {Schwenn}, R. 1998, Annales Geophysicae, 16, 1, \dodoi{10.1007/s00585-997-0001-x}

\bibitem[{{Burlaga} {et~al.}(1981){Burlaga}, {Sittler}, {Mariani}, \& {Schwenn}}]{burlaga1981}
{Burlaga}, L., {Sittler}, E., {Mariani}, F., \& {Schwenn}, R. 1981, \jgr, 86, 6673, \dodoi{10.1029/JA086iA08p06673}

\bibitem[{{Davies} {et~al.}(2021){Davies}, {M{\"o}stl}, {Owens}, {Weiss}, {Amerstorfer}, {Hinterreiter}, {Bauer}, {Bailey}, {Reiss}, {Forsyth}, {Horbury}, {O'Brien}, {Evans}, {Angelini}, {Heyner}, {Richter}, {Auster}, {Magnes}, {Baumjohann}, {Fischer}, {Barnes}, {Davies}, \& {Harrison}}]{davies2021}
{Davies}, E.~E., {M{\"o}stl}, C., {Owens}, M.~J., {et~al.} 2021, \aap, 656, A2, \dodoi{10.1051/0004-6361/202040113}

\bibitem[{{DeForest} {et~al.}(2013){DeForest}, {Howard}, \& {McComas}}]{deforest2013}
{DeForest}, C.~E., {Howard}, T.~A., \& {McComas}, D.~J. 2013, \apj, 769, 43, \dodoi{10.1088/0004-637X/769/1/43}

\bibitem[{{D{\'e}moulin} {et~al.}(2019){D{\'e}moulin}, {Dasso}, {Janvier}, \& {Lanabere}}]{demoulin2019}
{D{\'e}moulin}, P., {Dasso}, S., {Janvier}, M., \& {Lanabere}, V. 2019, \solphys, 294, 172, \dodoi{10.1007/s11207-019-1564-x}

\bibitem[{D'haeseleer {et~al.}(1991)D'haeseleer, Hitchon, Shohet, Callen, \& Kerst}]{d'haeseleer1991}
D'haeseleer, W.~D., Hitchon, W. N.~G., Shohet, J.~L., Callen, J.~D., \& Kerst, D.~W. 1991, Flux coordinates and magnetic field structure A guide to a fundamental tool of plasma theory (Germany: Springer).
\newblock \url{http://inis.iaea.org/search/search.aspx?orig_q=RN:23003745}

\bibitem[{{Fargette} {et~al.}(2020){Fargette}, {Lavraud}, {{\O}ieroset}, {Phan}, {Toledo-Redondo}, {Kieokaew}, {Jacquey}, {Fuselier}, {Trattner}, {Petrinec}, {Hasegawa}, {Garnier}, {G{\'e}not}, {Lenouvel}, {Fadanelli}, {Penou}, {Sauvaud}, {Avanov}, {Burch}, {Chandler}, {Coffey}, {Dorelli}, {Eastwood}, {Farrugia}, {Gershman}, {Giles}, {Grigorenko}, {Moore}, {Paterson}, {Pollock}, {Saito}, {Schiff}, \& {Smith}}]{fargette2020}
{Fargette}, N., {Lavraud}, B., {{\O}ieroset}, M., {et~al.} 2020, \grl, 47, e86726, \dodoi{10.1029/2019GL086726}

\bibitem[{{Galvin} {et~al.}(2008){Galvin}, {Kistler}, {Popecki}, {Farrugia}, {Simunac}, {Ellis}, {M{\"o}bius}, {Lee}, {Boehm}, {Carroll}, {Crawshaw}, {Conti}, {Demaine}, {Ellis}, {Gaidos}, {Googins}, {Granoff}, {Gustafson}, {Heirtzler}, {King}, {Knauss}, {Levasseur}, {Longworth}, {Singer}, {Turco}, {Vachon}, {Vosbury}, {Widholm}, {Blush}, {Karrer}, {Bochsler}, {Daoudi}, {Etter}, {Fischer}, {Jost}, {Opitz}, {Sigrist}, {Wurz}, {Klecker}, {Ertl}, {Seidenschwang}, {Wimmer-Schweingruber}, {Koeten}, {Thompson}, \& {Steinfeld}}]{galvin2008}
{Galvin}, A.~B., {Kistler}, L.~M., {Popecki}, M.~A., {et~al.} 2008, \ssr, 136, 437, \dodoi{10.1007/s11214-007-9296-x}

\bibitem[{{Gold} \& {Hoyle}(1960)}]{gold1960}
{Gold}, T., \& {Hoyle}, F. 1960, \mnras, 120, 89, \dodoi{10.1093/mnras/120.2.89}

\bibitem[{Hamada(1962)}]{hamada1962}
Hamada, S. 1962, Nuclear Fusion, 2, 23 .
\newblock \url{https://api.semanticscholar.org/CorpusID:121236776}

\bibitem[{{Isavnin}(2016)}]{isavnin2016}
{Isavnin}, A. 2016, \apj, 833, 267, \dodoi{10.3847/1538-4357/833/2/267}

\bibitem[{{Janvier} {et~al.}(2013){Janvier}, {D{\'e}moulin}, \& {Dasso}}]{janvier2013}
{Janvier}, M., {D{\'e}moulin}, P., \& {Dasso}, S. 2013, \aap, 556, A50, \dodoi{10.1051/0004-6361/201321442}

\bibitem[{{Jasinski} {et~al.}(2021){Jasinski}, {Akhavan-Tafti}, {Sun}, {Slavin}, {Coates}, {Fuselier}, {Sergis}, \& {Murphy}}]{jasinski2021}
{Jasinski}, J.~M., {Akhavan-Tafti}, M., {Sun}, W., {et~al.} 2021, Journal of Geophysical Research (Space Physics), 126, e28786, \dodoi{10.1029/2020JA028786}

\bibitem[{{Lepping} {et~al.}(1990){Lepping}, {Jones}, \& {Burlaga}}]{lepping1990}
{Lepping}, R.~P., {Jones}, J.~A., \& {Burlaga}, L.~F. 1990, \jgr, 95, 11957, \dodoi{10.1029/JA095iA08p11957}

\bibitem[{{Lepping} {et~al.}(1995){Lepping}, {Ac{\~{u}}na}, {Burlaga}, {Farrell}, {Slavin}, {Schatten}, {Mariani}, {Ness}, {Neubauer}, {Whang}, {Byrnes}, {Kennon}, {Panetta}, {Scheifele}, \& {Worley}}]{lepping1995}
{Lepping}, R.~P., {Ac{\~{u}}na}, M.~H., {Burlaga}, L.~F., {et~al.} 1995, \ssr, 71, 207, \dodoi{10.1007/BF00751330}

\bibitem[{{Lugaz} {et~al.}(2023){Lugaz}, {Lee}, {Jian}, {Allen}, {Al-Haddad}, {Winslow}, {Lillis}, {Moestl}, {Zhuang}, {Palmerio}, {Lynch}, {Scolini}, {Davies}, {Regnault}, {Nieves-Chinchilla}, \& {Farrugia}}]{lugaz2023}
{Lugaz}, N., {Lee}, C.~O., {Jian}, L.~K., {et~al.} 2023, in Bulletin of the American Astronomical Society, Vol.~55, 249, \dodoi{10.3847/25c2cfeb.ac2c2454}

\bibitem[{{Lugaz} {et~al.}(2024){Lugaz}, {Zhuang}, {Scolini}, {Al-Haddad}, {Farrugia}, {Winslow}, {Regnault}, {M{\"o}stl}, {Davies}, \& {Galvin}}]{lugaz2024}
{Lugaz}, N., {Zhuang}, B., {Scolini}, C., {et~al.} 2024, \apj, 962, 193, \dodoi{10.3847/1538-4357/ad17b9}

\bibitem[{{Lundquist}(1950)}]{lundquist1950}
{Lundquist}, S. 1950, Ark. Fys., 2, 361.
\newblock \url{https://ci.nii.ac.jp/naid/10003639556/en/}

\bibitem[{{Lynch} {et~al.}(2022){Lynch}, {Al-Haddad}, {Yu}, {Palmerio}, \& {Lugaz}}]{lynch2022}
{Lynch}, B.~J., {Al-Haddad}, N., {Yu}, W., {Palmerio}, E., \& {Lugaz}, N. 2022, Advances in Space Research, 70, 1614, \dodoi{10.1016/j.asr.2022.05.004}

\bibitem[{{Marubashi} \& {Lepping}(2007)}]{marubashi2007}
{Marubashi}, K., \& {Lepping}, R.~P. 2007, Annales Geophysicae, 25, 2453, \dodoi{10.5194/angeo-25-2453-2007}

\bibitem[{{Moldwin} {et~al.}(2000){Moldwin}, {Ford}, {Lepping}, {Slavin}, \& {Szabo}}]{moldwin2000}
{Moldwin}, M.~B., {Ford}, S., {Lepping}, R., {Slavin}, J., \& {Szabo}, A. 2000, \grl, 27, 57, \dodoi{10.1029/1999GL010724}

\bibitem[{{M{\"o}stl} {et~al.}(2020){M{\"o}stl}, {Weiss}, {Bailey}, {Reiss}, {Amerstorfer}, {Hinterreiter}, {Bauer}, {McIntosh}, {Lugaz}, \& {Stansby}}]{moestl2020}
{M{\"o}stl}, C., {Weiss}, A.~J., {Bailey}, R.~L., {et~al.} 2020, \apj, 903, 92, \dodoi{10.3847/1538-4357/abb9a1}

\bibitem[{{M{\"o}stl} {et~al.}(2022){M{\"o}stl}, {Weiss}, {Reiss}, {Amerstorfer}, {Bailey}, {Hinterreiter}, {Bauer}, {Barnes}, {Davies}, {Harrison}, {Freiherr von Forstner}, {Davies}, {Heyner}, {Horbury}, \& {Bale}}]{moestl2022}
{M{\"o}stl}, C., {Weiss}, A.~J., {Reiss}, M.~A., {et~al.} 2022, \apjl, 924, L6, \dodoi{10.3847/2041-8213/ac42d0}

\bibitem[{{Nieves-Chinchilla} {et~al.}(2023){Nieves-Chinchilla}, {Hidalgo}, \& {Cremades}}]{nieves2023}
{Nieves-Chinchilla}, T., {Hidalgo}, M.~A., \& {Cremades}, H. 2023, \apj, 947, 79, \dodoi{10.3847/1538-4357/acb3c1}

\bibitem[{{Nieves-Chinchilla} {et~al.}(2018){Nieves-Chinchilla}, {Linton}, {Hidalgo}, \& {Vourlidas}}]{nieves2018b}
{Nieves-Chinchilla}, T., {Linton}, M.~G., {Hidalgo}, M.~A., \& {Vourlidas}, A. 2018, \apj, 861, 139, \dodoi{10.3847/1538-4357/aac951}

\bibitem[{{Ogilvie} {et~al.}(1995){Ogilvie}, {Chornay}, {Fritzenreiter}, {Hunsaker}, {Keller}, {Lobell}, {Miller}, {Scudder}, {Sittler}, {Torbert}, {Bodet}, {Needell}, {Lazarus}, {Steinberg}, {Tappan}, {Mavretic}, \& {Gergin}}]{ogilvie1995}
{Ogilvie}, K.~W., {Chornay}, D.~J., {Fritzenreiter}, R.~J., {et~al.} 1995, \ssr, 71, 55, \dodoi{10.1007/BF00751326}

\bibitem[{{Owens} {et~al.}(2006){Owens}, {Merkin}, \& {Riley}}]{owens2006}
{Owens}, M.~J., {Merkin}, V.~G., \& {Riley}, P. 2006, Journal of Geophysical Research (Space Physics), 111, A03104, \dodoi{10.1029/2005JA011460}

\bibitem[{{Pal} {et~al.}(2023){Pal}, {Balmaceda}, {Weiss}, {Nieves-Chinchilla}, {Carcaboso}, {Kilpua}, \& {M{\"o}stl}}]{pal2023}
{Pal}, S., {Balmaceda}, L., {Weiss}, A.~J., {et~al.} 2023, Frontiers in Astronomy and Space Sciences, 10, 1195805, \dodoi{10.3389/fspas.2023.1195805}

\bibitem[{Richardson \& Cane(2024)}]{richardson_data}
Richardson, I., \& Cane, H. 2024, {Near-Earth Interplanetary Coronal Mass Ejections Since January 1996}, V2,  Harvard Dataverse, \dodoi{10.7910/DVN/C2MHTH}

\bibitem[{{Riley} \& {Crooker}(2004)}]{riley2004}
{Riley}, P., \& {Crooker}, N.~U. 2004, \apj, 600, 1035, \dodoi{10.1086/379974}

\bibitem[{{Rodr{\'\i}guez} {et~al.}(2021){Rodr{\'\i}guez}, {Sengupta}, \& {Bhattacharjee}}]{rodriguez2021}
{Rodr{\'\i}guez}, E., {Sengupta}, W., \& {Bhattacharjee}, A. 2021, Physics of Plasmas, 28, 092510, \dodoi{10.1063/5.0060115}

\bibitem[{Salman {et~al.}(2024)Salman, Nieves-Chinchilla, Jian, Lugaz, Carcaboso, Davies, \& Collado-Vega}]{salman2024}
Salman, T.~M., Nieves-Chinchilla, T., Jian, L.~K., {et~al.} 2024, A Survey of Coronal Mass Ejections Measured In Situ by Parker Solar Probe During 2018-2022.
\newblock \doarXiv{2403.02594}

\bibitem[{{Slavin} {et~al.}(2003){Slavin}, {Lepping}, {Gjerloev}, {Fairfield}, {Hesse}, {Owen}, {Moldwin}, {Nagai}, {Ieda}, \& {Mukai}}]{slavin2003a}
{Slavin}, J.~A., {Lepping}, R.~P., {Gjerloev}, J., {et~al.} 2003, Journal of Geophysical Research (Space Physics), 108, 1015, \dodoi{10.1029/2002JA009557}

\bibitem[{{Solov'ev} \& {Kirichek}(2021)}]{solovev2021}
{Solov'ev}, A.~A., \& {Kirichek}, E.~A. 2021, \mnras, 505, 4406, \dodoi{10.1093/mnras/stab1565}

\bibitem[{{Song} {et~al.}(2020){Song}, {Zhang}, {Cheng}, {Li}, {Hu}, {Li}, {Chen}, {Zheng}, \& {Chen}}]{song2020}
{Song}, H.~Q., {Zhang}, J., {Cheng}, X., {et~al.} 2020, \apjl, 901, L21, \dodoi{10.3847/2041-8213/abb6ec}

\bibitem[{{Thernisien} {et~al.}(2009){Thernisien}, {Vourlidas}, \& {Howard}}]{thernisien2009}
{Thernisien}, A., {Vourlidas}, A., \& {Howard}, R.~A. 2009, \solphys, 256, 111, \dodoi{10.1007/s11207-009-9346-5}

\bibitem[{{Thernisien} {et~al.}(2006){Thernisien}, {Howard}, \& {Vourlidas}}]{thernisien2006}
{Thernisien}, A.~F.~R., {Howard}, R.~A., \& {Vourlidas}, A. 2006, \apj, 652, 763, \dodoi{10.1086/508254}

\bibitem[{{Titov} \& {D{\'e}moulin}(1999)}]{titov1999}
{Titov}, V.~S., \& {D{\'e}moulin}, P. 1999, \aap, 351, 707

\bibitem[{{T{\"o}r{\"o}k} \& {Kliem}(2005)}]{toeroek2005}
{T{\"o}r{\"o}k}, T., \& {Kliem}, B. 2005, \apjl, 630, L97, \dodoi{10.1086/462412}

\bibitem[{{Vandas} \& {Romashets}(2017{\natexlab{a}})}]{vandas2017b}
{Vandas}, M., \& {Romashets}, E. 2017{\natexlab{a}}, \aap, 608, A118, \dodoi{10.1051/0004-6361/201731412}

\bibitem[{{Vandas} \& {Romashets}(2017{\natexlab{b}})}]{vandas2017a}
---. 2017{\natexlab{b}}, \solphys, 292, 129, \dodoi{10.1007/s11207-017-1149-5}

\bibitem[{{Verbeke} {et~al.}(2023){Verbeke}, {Mays}, {Kay}, {Riley}, {Palmerio}, {Dumbovi{\'c}}, {Mierla}, {Scolini}, {Temmer}, {Paouris}, {Balmaceda}, {Cremades}, \& {Hinterreiter}}]{verbeke2023}
{Verbeke}, C., {Mays}, M.~L., {Kay}, C., {et~al.} 2023, Advances in Space Research, 72, 5243, \dodoi{10.1016/j.asr.2022.08.056}

\bibitem[{{Vourlidas}(2014)}]{vourlidas2014}
{Vourlidas}, A. 2014, Plasma Physics and Controlled Fusion, 56, 064001, \dodoi{10.1088/0741-3335/56/6/064001}

\bibitem[{{Wang} \& {Liu}(2019)}]{wang2019}
{Wang}, H., \& {Liu}, C. 2019, Frontiers in Astronomy and Space Sciences, 6, 18, \dodoi{10.3389/fspas.2019.00018}

\bibitem[{{Weiss} {et~al.}(2021{\natexlab{a}}){Weiss}, {M{\"o}stl}, {Amerstorfer}, {Bailey}, {Reiss}, {Hinterreiter}, {Amerstorfer}, \& {Bauer}}]{weiss2021a}
{Weiss}, A.~J., {M{\"o}stl}, C., {Amerstorfer}, T., {et~al.} 2021{\natexlab{a}}, \apjs, 252, 9, \dodoi{10.3847/1538-4365/abc9bd}

\bibitem[{{Weiss} {et~al.}(2022){Weiss}, {Nieves-Chinchilla}, {M{\"o}stl}, {Reiss}, {Amerstorfer}, \& {Bailey}}]{weiss2022}
{Weiss}, A.~J., {Nieves-Chinchilla}, T., {M{\"o}stl}, C., {et~al.} 2022, Journal of Geophysical Research (Space Physics), 127, e2022JA030898, \dodoi{10.1029/2022JA030898}

\bibitem[{{Weiss} {et~al.}(2021{\natexlab{b}}){Weiss}, {M{\"o}stl}, {Davies}, {Amerstorfer}, {Bauer}, {Hinterreiter}, {Reiss}, {Bailey}, {Horbury}, {O'Brien}, {Evans}, {Angelini}, {Heyner}, {Richter}, {Auster}, {Magnes}, {Fischer}, \& {Baumjohann}}]{weiss2021b}
{Weiss}, A.~J., {M{\"o}stl}, C., {Davies}, E.~E., {et~al.} 2021{\natexlab{b}}, \aap, 656, A13, \dodoi{10.1051/0004-6361/202140919}

\bibitem[{{Zurbuchen} \& {Richardson}(2006)}]{zurbuchen2006}
{Zurbuchen}, T.~H., \& {Richardson}, I.~G. 2006, \ssr, 123, 31, \dodoi{10.1007/s11214-006-9010-4}

\end{thebibliography}
\bibliographystyle{aasjournal}

\end{document}